\documentclass[twocolumn]{aastex631}
\hypersetup{linkcolor=red,citecolor=blue,filecolor=cyan,urlcolor=magenta}

\usepackage{amsmath}
\usepackage{comment}
\usepackage[british, calc]{datetime2}
\DTMlangsetup[en-GB]{ord=raise}

\shorttitle{The Type Ibn SN SN\,2019kbj}
\shortauthors{Ben-Ami et al.}

\begin{document}

\title{The Type Ibn Supernova 2019kbj - Indications for
Diversity in Type Ibn Supernova Progenitors}
\DTMsetstyle{default}\

\author[0000-0001-7513-6701]{Tom Ben-Ami}
\affiliation{Cavendish Laboratory, University of Cambridge, J. J. Thomson Avenue, Cambridge, CB3 0HE, UK}
\affiliation{The School of Physics and Astronomy, Tel Aviv University, Tel Aviv 69978, Israel}

\author[0000-0001-7090-4898]{Iair Arcavi}
\affiliation{The School of Physics and Astronomy, Tel Aviv University, Tel Aviv 69978, Israel}
\affiliation{CIFAR Azrieli Global Scholars Program, CIFAR, Toronto, Canada}

\author{Megan Newsome}
\affiliation{Department of Physics, University of California, Santa Barbara, Santa Barbara, CA 93106-9530, USA}
\affiliation{Las Cumbres Observatory, 6740 Cortona Dr Ste 102, Goleta, CA 93117-5575, USA}

\author{Joseph Farah}
\affiliation{Department of Physics, University of California, Santa Barbara, CA 93106-9530, USA}
\affiliation{Las Cumbres Observatory, 6740 Cortona Dr Ste 102, Goleta, CA 93117-5575, USA}

\author[0000-0002-7472-1279]{Craig Pellegrino}
\affiliation{Department of Physics, University of California, Santa Barbara, CA 93106-9530, USA}
\affiliation{Las Cumbres Observatory, 6740 Cortona Dr Ste 102, Goleta, CA 93117-5575, USA}

\author{Giacomo Terreran}
\affiliation{Department of Physics, University of California, Santa Barbara, CA 93106-9530, USA}
\affiliation{Las Cumbres Observatory, 6740 Cortona Dr Ste 102, Goleta, CA 93117-5575, USA}

\author{Jamison Burke}
\affiliation{Department of Physics, University of California, Santa Barbara, CA 93106-9530, USA}
\affiliation{Las Cumbres Observatory, 6740 Cortona Dr Ste 102, Goleta, CA 93117-5575, USA}

\author[0000-0002-0832-2974]{Griffin Hosseinzadeh}
\affiliation{Steward Observatory, University of Arizona, 933 North Cherry Avenue, Tucson, AZ 85721-0065, USA}

\author{Curtis McCully}
\affiliation{Las Cumbres Observatory, 6740 Cortona Dr Ste 102, Goleta, CA 93117-5575, USA}

\author[0000-0002-1125-9187]{Daichi Hiramatsu}
\affiliation{Center for Astrophysics, Harvard \& Smithsonian, 60 Garden Street, Cambridge, Ma 02138, USA}
\affiliation{The NSF AI Institute for Artificial Intelligence and Fundamental Interactions, USA}

\author[0000-0003-0209-9246]{Estefania Padilla Gonzalez}
\affiliation{Department of Physics, University of California, Santa Barbara, CA 93106-9530, USA}
\affiliation{Las Cumbres Observatory, 6740 Cortona Dr Ste 102, Goleta, CA 93117-5575, USA}

\author{D. Andrew Howell}
\affiliation{Department of Physics, University of California, Santa Barbara, CA 93106-9530, USA}
\affiliation{Las Cumbres Observatory, 6740 Cortona Dr Ste 102, Goleta, CA 93117-5575, USA}

\correspondingauthor{Iair Arcavi}
\email{arcavi@tauex.tau.ac.il}

\begin{abstract}
Type Ibn supernovae (SNe) are a rare class of stellar explosions whose progenitor systems are not yet well determined. We present and analyze observations of the Type Ibn SN\,2019kbj, and model its light curve in order to constrain its progenitor and explosion parameters. SN\,2019kbj shows roughly constant temperature during the first month after peak, indicating a power source (likely interaction with circumstellar material) that keeps the continuum emission hot at $\sim15,000$\,K. Indeed, we find that the radioactive decay of $^{56}$Ni is disfavored as the sole power source of the bolometric light curve. A radioactive decay + circumstellar-material (CSM) interaction model, on the other hand, does reproduce the bolometric emission well. The fits prefer a uniform-density CSM shell rather than CSM due to a steady mass-loss wind, similar to what is seen in other Type Ibn SNe. The uniform-density CSM shell model requires $\sim0.1 M_\odot$ of $^{56}$Ni and $\sim1 M_\odot$ total ejecta mass to reproduce the light curve. SN\,2019kbj differs in this manner from another Type Ibn SN with derived physical parameters, SN\,2019uo, for which an order of magnitude lower $^{56}$Ni mass and larger ejecta mass were derived. This points toward a possible diversity in SN Ibn progenitor systems and explosions. 
\end{abstract}

\section{Introduction} \label{sec:intro}

Type Ibn supernovae (SNe) are a rare class of stellar explosions characterized by a lack of hydrogen lines and the presence of narrow \ion{He}{1} emission lines in their spectra \citep{pastorello2007}. These events are thought to be SNe strongly interacting with H-poor, helium-rich circumstellar material \citep[CSM; e.g.][and references therein]{Smith2016}. 
Only a few dozen of such events are known \citep[see][for recent compilations]{pastorello16,Hosseinzadeh_2017}, and their progenitor systems remain a mystery.

The H-rich analogs of Type Ibn events, (i.e. explosions interacting with a H-rich CSM), known as Type IIn SNe, show slowly evolving and diverse light curves \citep[e.g.][]{Kiewe2012}. These traits are explained by the fact that CSM interaction injects extra luminosity through shocks, producing the observed prolonged emission, while diverse CSM density distributions produce the observed diversity in light-curve shapes. However, \cite{Hosseinzadeh_2017} showed that many type Ibn SN light curves are strikingly similar and rapidly evolving, in contrast to the expectations from CSM-interaction-powered emission (but see also outliers to this uniformity discussed in \citealt{pastorello16}). 

Even more puzzling is the discovery of a Type Ibn SN in a brightest cluster galaxy \citep{sanders13}, specifically in an environment with extremely low star formation, leading \cite{Hosseinzadeh2019} to conclude that some (if not all) Type Ibn SNe might not even be explosions of massive stars, as typically assumed.

Possible clues as to the progenitors of Type Ibn SNe can come from modeling their bolometric light curves. \cite{Gangopadhyay_2020} fit the bolometric light curve of the Type Ibn SN\,2019uo with the \cite{chatzopoulos12} model that includes luminosity from both $^{56}$Ni decay and CSM interaction (after disfavoring $^{56}$Ni decay as the sole power source). Their best fits require $\sim16M_{\odot}$ of ejecta and just $0.01M_{\odot}$ of $^{56}$Ni, with most of the luminosity at peak coming from interaction of the ejecta with a few tenths of a solar mass of CSM. They favor a uniform-density shell, rather than a steady wind, for the distribution of the CSM. \cite{Pellegrino2022}, on the other hand, find a much smaller ejecta mass ($\sim1M_{\odot}$) for the same event, while finding a similar $^{56}$Ni mass, using the same models.

Here we present observations of SN\,2019kbj, a well-observed member of the Type Ibn class, with multiband photometry and multiepoch spectroscopy. We analyze its light curve and spectra and model its bolometric light curve in a similar way to that of \citealt{Gangopadhyay_2020} for SN\,2019uo to deduce its physical parameters. With this analysis we aim to increase the sample of Type Ibn events with deduced physical parameters.  We assume the \texttt{Planck18} \citep{planck18} cosmology throughout.

\begin{deluxetable}{ccccc}[b]
\tablecaption{Photometry of SN\,2019kbj.} \label{tab:rawdat}
\tablehead{
 \colhead{MJD} & \colhead{Filter} & \colhead{Magnitude} & \colhead{Error} & \colhead{Source} 
}
\startdata
 58663.49 &    $c$ & $<$19.89 &     &  ATLAS \\
 58665.49 &    $o$ & 18.38 &  0.080 &  ATLAS \\
 58665.50 &    $o$ & 18.17 &  0.078 &  ATLAS \\
 58665.50 &    $o$ & 18.27 &  0.067 &  ATLAS \\
 58665.51 &    $o$ & 18.14 &  0.059 &  ATLAS \\
 58667.46 &    $c$ & 17.60 &  0.033 &  ATLAS \\
 58667.48 &    $c$ & 17.64 &  0.034 &  ATLAS \\
 58667.50 &    $c$ & 17.61 &  0.033 &  ATLAS \\
 58667.50 &    $c$ & 17.55 &  0.028 &  ATLAS \\
 58668.39 &    $B$ & 17.44 &  0.018 &  Las Cumbres \\
 58668.39 &    $B$ & 17.36 &  0.009 &  Las Cumbres \\
 58668.40 &    $V$ & 17.55 &  0.013 &  Las Cumbres \\
 58668.40 &    $V$ & 17.55 &  0.013 &  Las Cumbres \\
 58668.40 &    $g$ & 17.30 &  0.006 &  Las Cumbres \\
 58668.40 &    $g$ & 17.30 &  0.006 &  Las Cumbres \\
\enddata
\tablecomments{This table is published in its entirety in machine-readable format. A portion is shown here for guidance regarding its form and content.}
\end{deluxetable}

\section{Discovery and Classification} \label{sec:discover}

SN\,2019kbj was discovered on 2019 July 1 (UT used throughout) by the Asteroid Terrestrial-impact Last Alert System \citep[ATLAS;][]{Tonry2018} transient survey \citep{Smith2020} as ATLAS19ohl \citep{discovery}, at R.A. \texttt{01:00:39.619} and decl. \texttt{+19:37:03.5} (J2000)\footnote{The event was independently discovered on 2019 July 27 by the Panoramic Survey Telescope And Rapid Response System \citep[Pan-STARRS;][]{Chambers2016} as PS19dzw.}. A faint (absolute magnitude $\sim-17$) and blue host galaxy is seen in archival PS1 images \citep{Flewelling2020} at this position.

The event was initially classified on 2019 July 3 by \cite{classification} as a possible young Type II SN at a redshift of $z=0.048$, based on the strong blue continuum, narrow H emission, and possible early flash-spectroscopy features \citep[short-lived high-ionization emission lines indicative of a confined CSM; e.g.][]{Khazov2016}. However, it was later reclassified by \cite{Arcavi2022} as a Type Ibn SN based on narrow He I emission lines (and a lack of broad H features) seen in a spectrum taken one week later (the narrow H emission being attributed to the host galaxy rather than the SN). The redshift remained unrevised. 

\begin{deluxetable}{cccc}[b]
\tablecaption{Log of spectroscopic observations.}\label{tab:logs}
\tablehead{
\colhead{Date} & \colhead{MJD} & \colhead{Phase} & \colhead{Telescope}\\
\colhead{} & \colhead{} & \colhead{($days$)} & \colhead{}}
\startdata
    \DTMdate{2019-07-02} & 58666.68 & $-2.42$ & FTS 2m\\
    \DTMdate{2019-07-03} & 58667.49 & $-1.61$ & FTN 2m\\
    \DTMdate{2019-07-04} & 58668.43 & $-0.68$ & FTN 2m\\ 
    \DTMdate{2019-07-10} & 58674.43 & $5.32$ & FTN 2m\\ 
    \DTMdate{2019-07-13} & 58677.57 & $8.47$ & FTN 2m\\ 
    \DTMdate{2019-07-15} & 58679.52 & $10.42$ & FTN 2m\\ 
    \DTMdate{2019-07-18} & 58682.48 & $13.38$ & FTN 2m\\ 
    \DTMdate{2019-07-24} & 58688.56 & $19.45$ & FTN 2m\\ 
    \DTMdate{2019-07-28} & 58692.47 & $23.36$ & FTN 2m\\ 
    \DTMdate{2019-08-05} & 58700.52 & $31.41$ & FTN 2m\\ 
    \DTMdate{2019-08-09} & 58704.53 & $35.43$ & FTN 2m\\ 
    \DTMdate{2022-05-23} & 59722.52 & Host & FTN 2m\\
\enddata
\end{deluxetable}

\begin{figure*}
    \centering
    \includegraphics[width=0.9\textwidth]{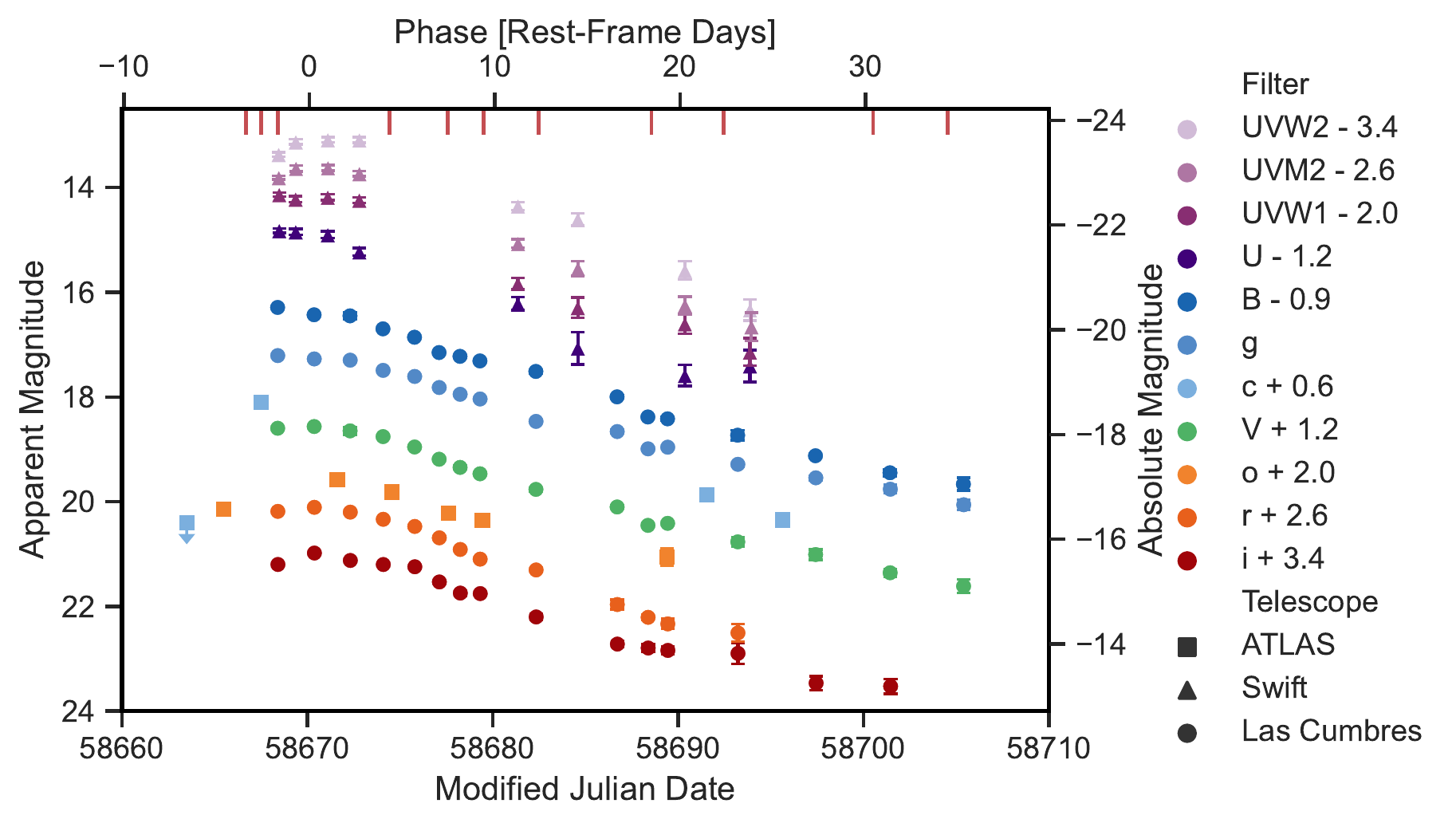}
    \caption{Extinction-corrected multiband light curve of SN\,2019kbj. Vertical red lines at the top indicate days when spectra were obtained. The arrow indicates the last nondetection $5\sigma$ limit.}
    \label{fig:phot1}
\end{figure*}

\section{Observations and Data Reduction} \label{sec:obs}

We obtained $BVgri$-band imaging of SN\,2019kbj with the Las Cumbres Observatory \citep{Brown2013} Sinistro cameras mounted on the network of 1-meter telescopes at the Cerro Tololo Inter-American Observatory (Chile), the South African Astronomical Observatory (South Africa), the Siding Spring Observatory (Australia), and the McDonald Observatory (United States), through the Global Supernova Project, from 2019 July 4 to 2019 September 20. Reference images were obtained on 2021 December 31, long after the SN faded. Standard image-reduction procedures were applied by the Las Cumbres Beautiful Algorithms to Normalize Zillions of Astronomical Images (BANZAI) pipeline\footnote{\url{https://github.com/LCOGT/banzai}} \citep{McCully2018}. We then performed image subtraction and point-spread function (PSF) fitting using the PyRAF-based \texttt{lcogtsnpipe}\footnote{\url{https://github.com/LCOGT/lcogtsnpipe}} pipeline \citep{valenti16}, which uses the High Order Transform of PSF ANd Template Subtraction \citep[HOTPANTS;][]{Becker2015} implementation of the \cite{Alard1998} algorithm. $BV$-band magnitudes are calibrated to the Vega system, and $gri$-band magnitudes to the AB system. 
We also obtained $c$- and $o$-band host-subtracted photometry of SN\,2019kbj from the ATLAS Forced Photometry Server\footnote{\url{https://fallingstar-data.com/forcedphot/}} \citep{Tonry2018,Smith2020}. We find the last pre-explosion ATLAS 5$\sigma$ nondetection limit to be on 2019 June 29 at a magnitude of $19.89$ in the $c$ band, constraining the explosion time to a window of only $2$ days between 2019 June 29 and 2019 July 1. 

We downloaded images of SN\,2019kbj taken by the Ultraviolet Optical Telescope (UVOT; \citealt{Roming05}) on board the {\it Neil Gehrels Swift Observatory} \citep{Gehrels04}, obtained under a Target of Opportunity request (PI: Hiramatsu), from the High Energy Astrophysics Science Archive Research Center (HEASARC)\footnote{\url{https://heasarc.gsfc.nasa.gov/}}. We performed aperture photometry with a 5$\arcsec$-radius circular region using the \texttt{uvotsource} package in HEAsoft v6.18, with version 20200925 of the calibration database (CALDB), following the standard guidelines from \cite{Brown09}. Host flux subtraction was performed using images taken on 2022 April 12 (PI: Grupe), long after the SN faded, following the prescriptions of \cite{Brown14}.

\begin{deluxetable}{lllllll}[b]
\tablecaption{Post-peak luminosity decline rates of SN\,2019kbj in magnitudes per day. These values are typical for Type Ibn SNe.}\label{tab:decline} 
\tablehead{
\colhead{} & \colhead{$B$} & \colhead{$g$} & \colhead{$V$} & \colhead{$o$} & \colhead{$r$} & \colhead{$i$}}
\startdata
    Decline rate & 0.099 & 0.078 & 0.092  & 0.083 & 0.12 & 0.091\\ 
    Error & 0.003 & 0.031 & 0.003 & 0.006 & 0.004 & 0.005
\enddata
\end{deluxetable}

\begin{figure*}
    \centering
    \includegraphics[width=0.9\textwidth]{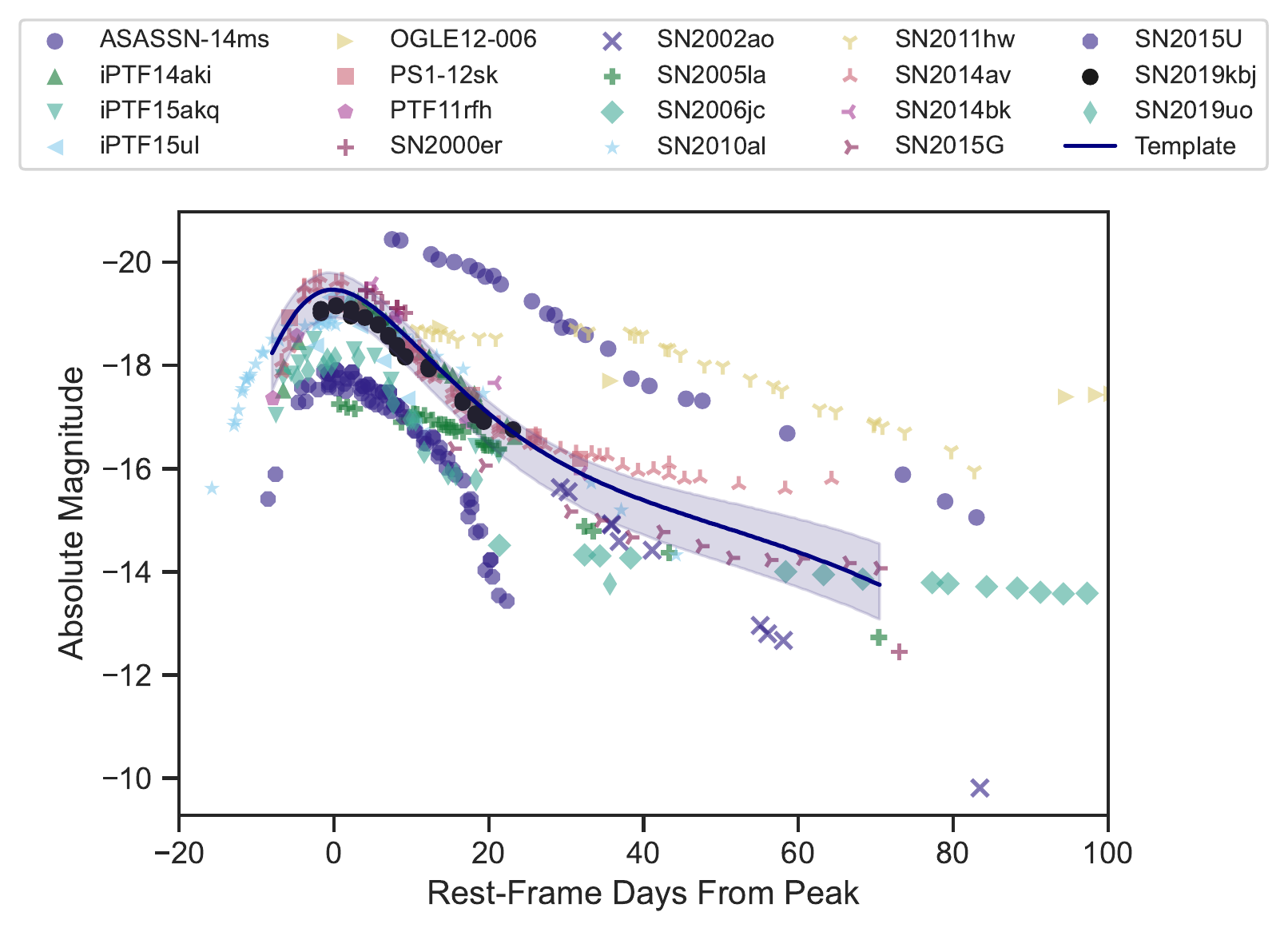}
    \caption{Absolute magnitude $r$-band light curve of SN\,2019kbj (black circles) compared to $r$ and $R$-band light curves of other Type Ibn SNe and the Type Ibn SN $r$-band template (shaded region) from \cite{Hosseinzadeh_2017}. SN\,2019kbj fits well within the population of Ibn SNe. The data for ASASSN-14ms are taken from \cite{wang21}, PTF11rfh, iPTF15ul, iPTF15akq, iPTF14aki and SN\,2015U from \cite{Hosseinzadeh_2017}, OGLE12-006 from \cite{pastorello15e}, PS1-12sk from \cite{sanders13}, SN\,2000er and SN\,2002ao from \cite{pastorello08}, SN\,2005la from \cite{pastorello08b}, SN\,2006jc from \cite{pastorello2007,pastorello08}, SN\,2010al and SN\,2011hw from \cite{pastorello15a}, SN2014av and SN\,2014bk from \cite{pastorello16}, SN\,2015G from \cite{foley15} and SN\,2019uo from \cite{Gangopadhyay_2020}.
    \label{fig:phot2}
    }
\end{figure*}

We correct all photometry for Milky Way extinction using the \cite{galacticrecal} calibrations of the \cite{galacticsurvey} maps, retrieved via the NASA/IPAC Extragalactic Database (NED)\footnote{\url{https://ned.ipac.caltech.edu/}}. For the ATLAS $c$ and $o$ bands we use extinction data for the $g$ and $r$ bands, respectively. We neglect extinction in the SN host galaxy, as we find no evidence for strong \ion{Na}{1} D absorption in a spectrum taken of the host (see below). Our photometry is presented in Table \ref{tab:rawdat} and in Figures \ref{fig:phot1}--\ref{fig:phot3}.

We obtained 12 spectroscopic observations with the Las Cumbres Observatory Floyds spectrographs mounted on the 2-meter Faulkes Telescope North (FTN) and South (FTS) at Haleakala (United States) and Siding Spring (Australia) observatories, respectively, through the Global Supernova Project. Spectra were obtained through a 2\arcsec\ slit placed on the SN along the parallactic angle \citep{Filippenko_1982}. One-dimensional spectra were extracted, and flux and wavelength calibrated using the \texttt{floyds\_pipeline}\footnote{\url{https://github.com/LCOGT/floyds_pipeline}} \citep{Valenti2013}. One of the spectra is of the host galaxy, obtained long after the SN faded. A log of the spectroscopic measurements is given in Table \ref{tab:logs}. All SN spectra are presented in Figure \ref{fig:spec1}. The host spectrum is presented in Figure \ref{fig:galactic_spec}.

\section{Photometric Analysis}
\label{sec:lc}

The multiband light curve of SN\,2019kbj is shown in Figure \ref{fig:phot1}. Using a parabolic fit to the $r$-band data around peak (from MJD 58668.4 to 58677.1), we determine the peak date to be MJD $58670.1 \pm 0.26$, with an apparent peak magnitude of $17.67 \pm 0.24$, corresponding to an absolute peak magnitude of $-18.99\mathbf{\pm 0.24}$ (errors are from the parabolic fit).

We calculate the post-peak decline rate using a linear fit to the magnitudes between MJD $58670$ and $58700$ for each band (except the $c$ band for which there are not enough epochs). Our results are presented in Table \ref{tab:decline}. We find a decline rate in all bands similar to the typical $r$-band 0.1 mag~day\textsuperscript{-1} measured for Type Ibn SNe by \cite{Hosseinzadeh_2017}. 

Comparing the $r$-band light curve of SN\,2019kbj to those of other Type Ibn SNe (Fig. \ref{fig:phot2}), we find that it is rather typical and fits well within the template of \cite{Hosseinzadeh_2017} around peak. SN\,2019kbj shows excess emission compared to the template starting at around 20 days after peak, perhaps due to a larger amount of $^{56}$Ni compared to other events (see below). 

\begin{figure}
    \includegraphics[width=0.48\textwidth]{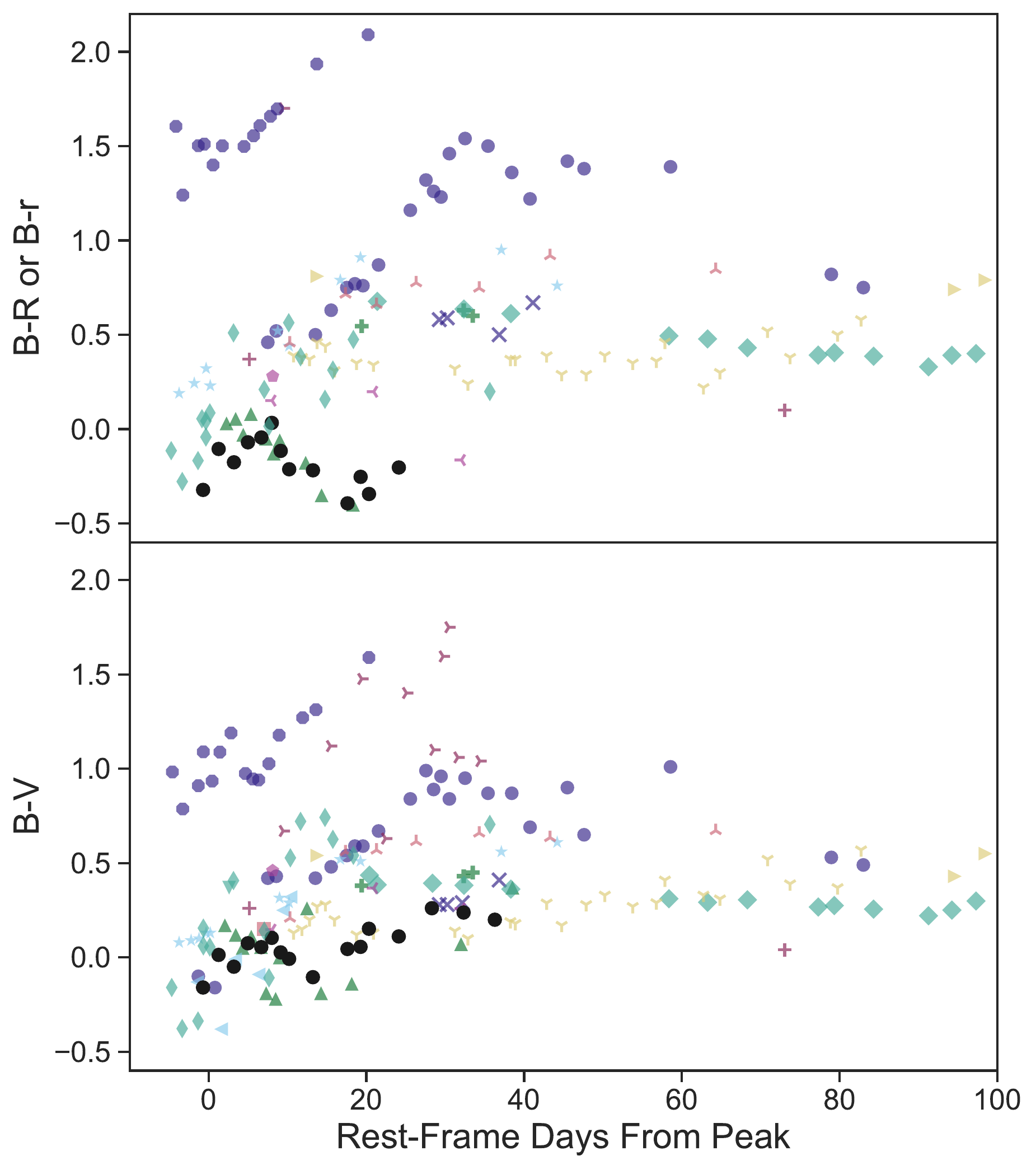}
    \caption{The color evolution of SN\,2019kbj (black circles) compared to other Type Ibn SNe. SN\,2019kbj shows constant colors, as do some other Type Ibn SNe. Colors, symbols and data sources are as in Figure \ref{fig:phot2}.}
    \label{fig:phot3}
\end{figure}

The color evolution of SN\,2019kbj is shown in Figure \ref{fig:phot3}. Both its $B-r$ and $B-V$ colors are roughly constant, as seen also in other Type Ibn SNe. SN\,2019kbj is one of the bluest Ibn's in the sample. 

\section{Spectroscopic Analysis} \label{sec:spec}

\begin{figure*}
    \centering
    \includegraphics[width=0.5\textwidth]{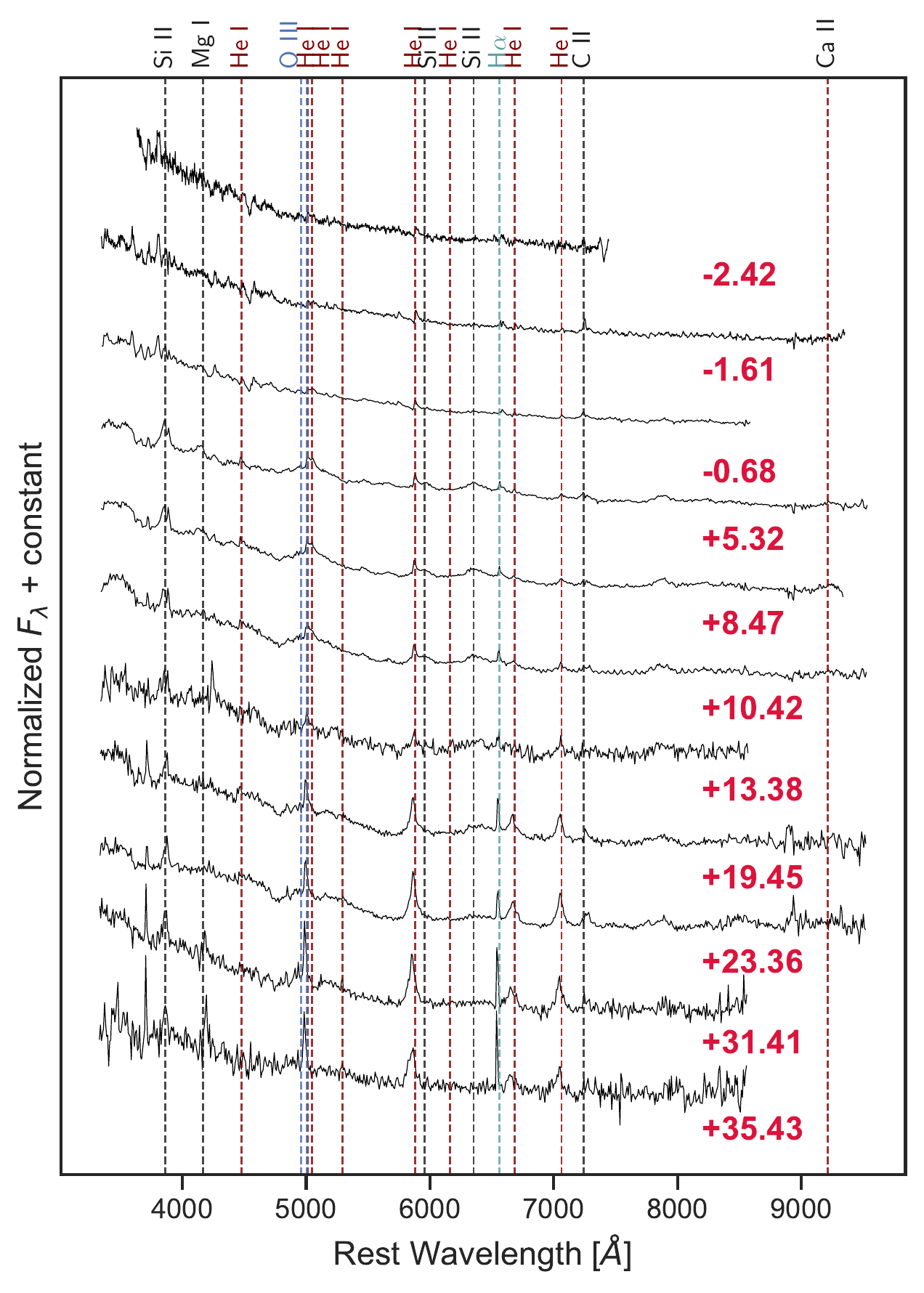}\,\includegraphics[width=0.5\textwidth]{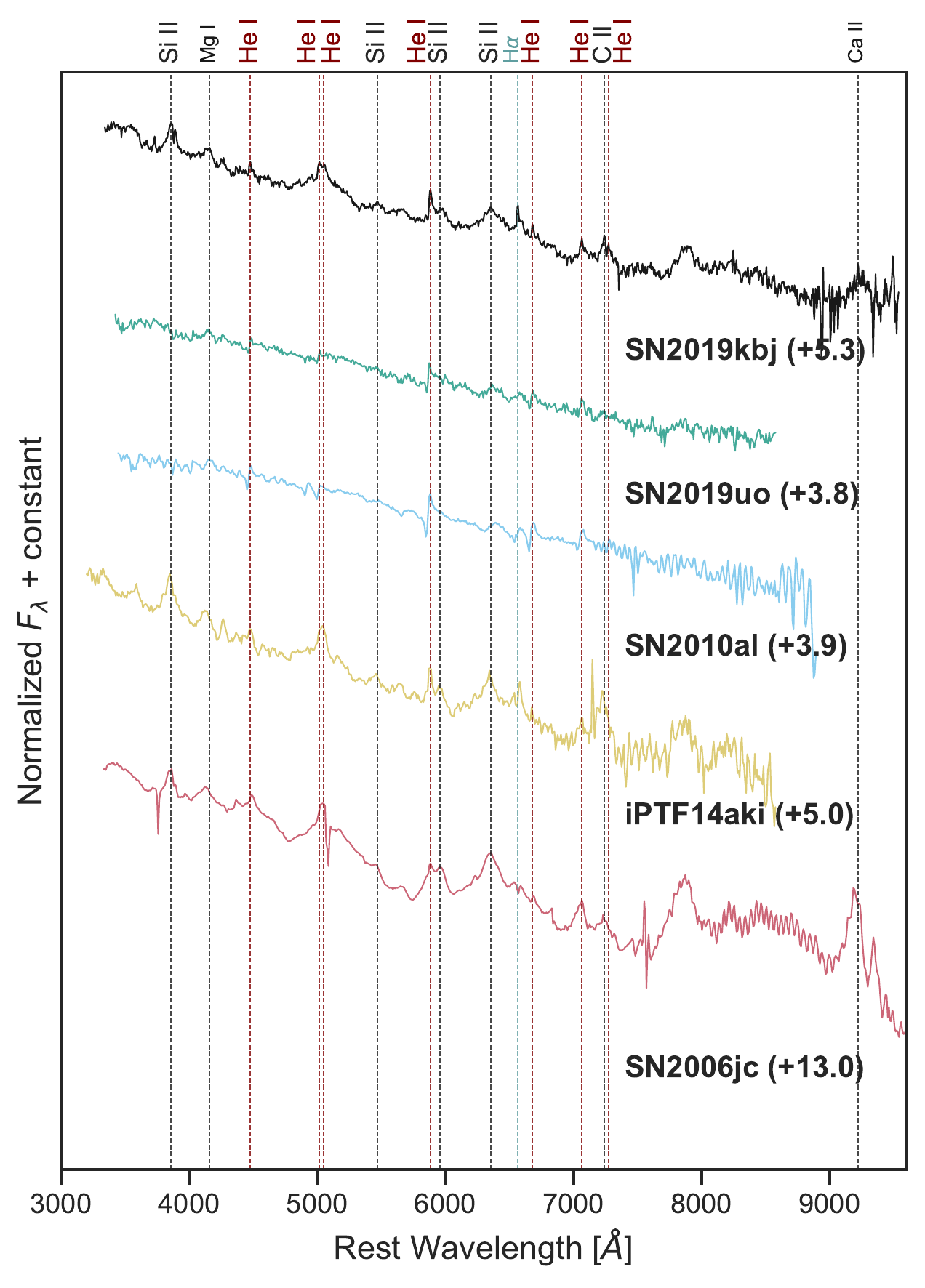}
    \caption{Left: The spectral evolution of SN\,2019kbj. Prominent spectral lines and days relative to peak luminosity are noted. Right: The spectrum of SN\,2019kbj 5.3 days after peak compared to other Type Ibn SNe at similar phases (noted in days relative to peak).}
    \label{fig:spec1}
\end{figure*}

\begin{figure*}
    \centering
    \renewcommand*{\arraystretch}{0.2}
    \begin{tabular}{*{5}{@{}c}@{}}
    \includegraphics[width=0.2\textwidth]{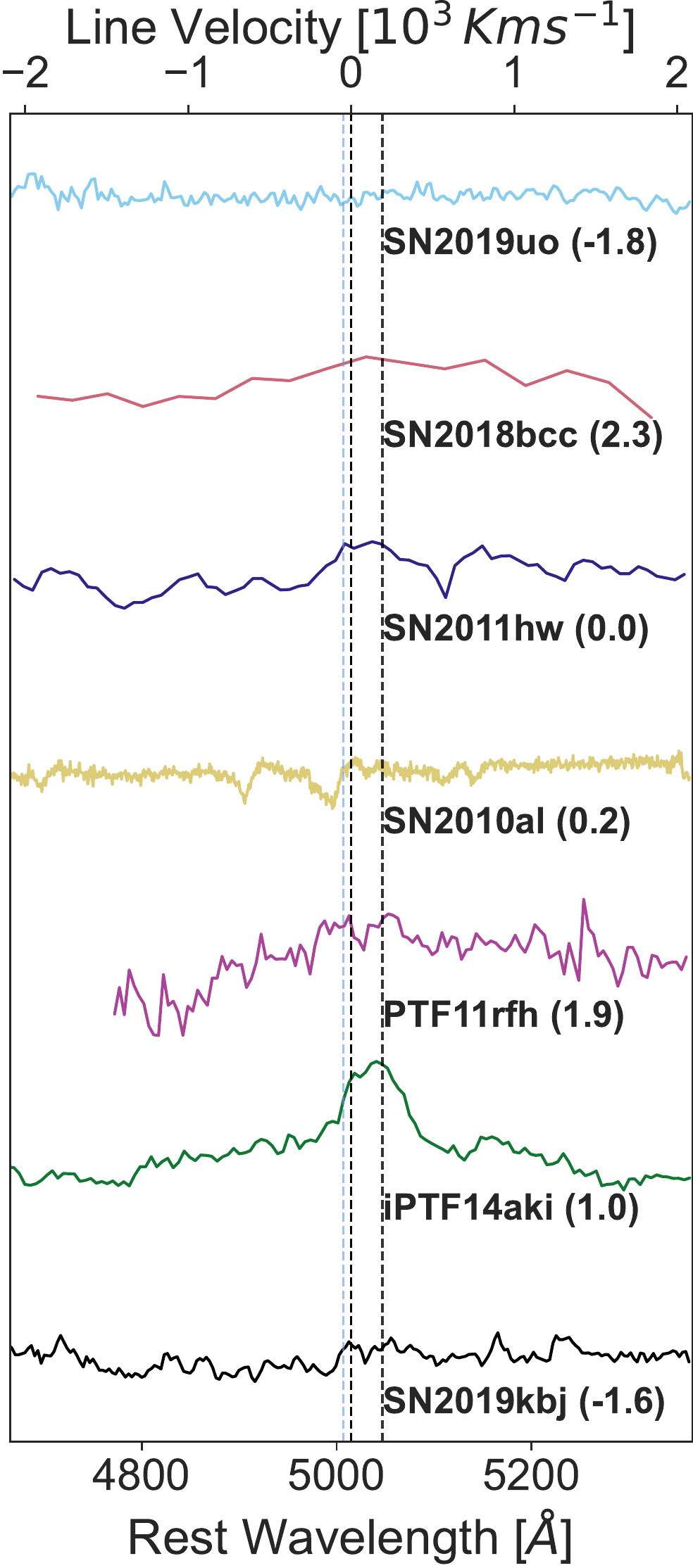} &
    \includegraphics[width=0.2\textwidth]{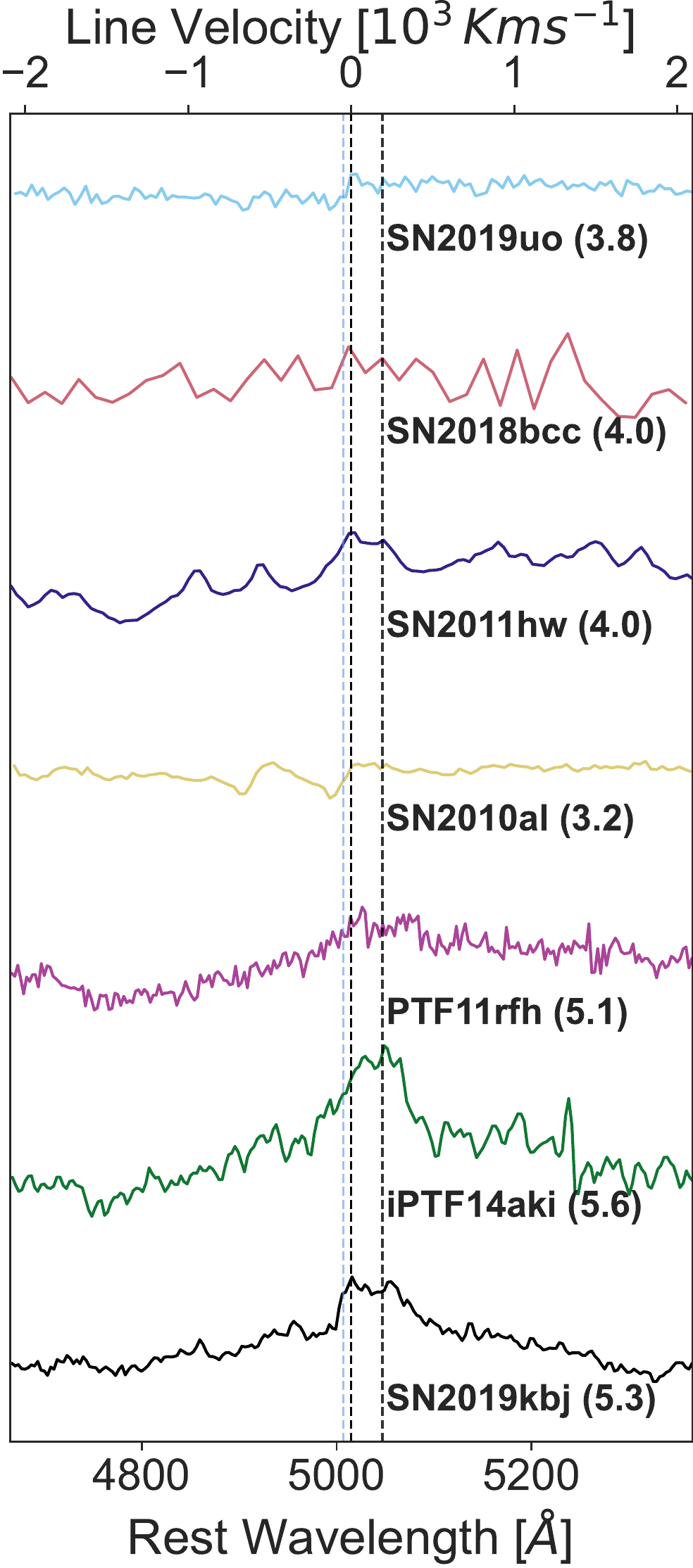} &
    \includegraphics[width=0.2\textwidth]{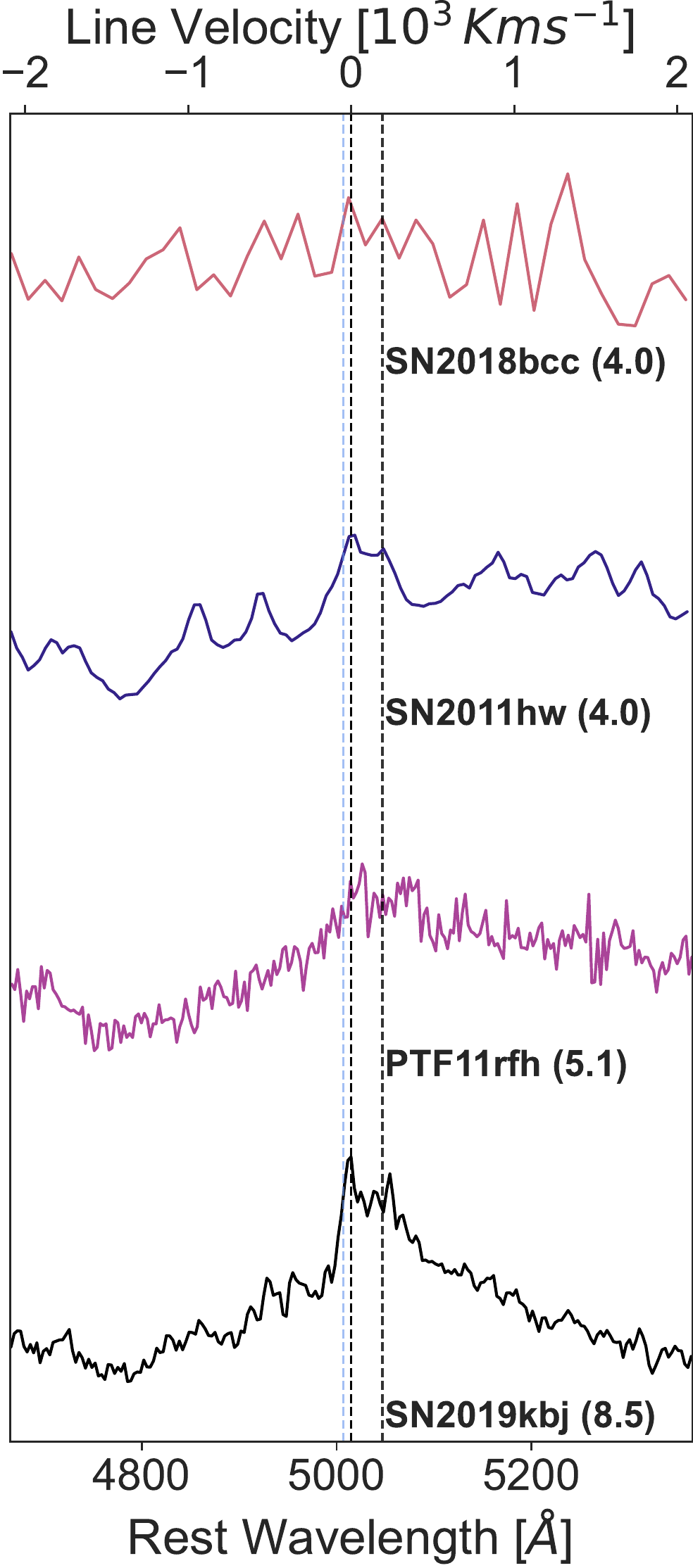} &
    \includegraphics[width=0.2\textwidth]{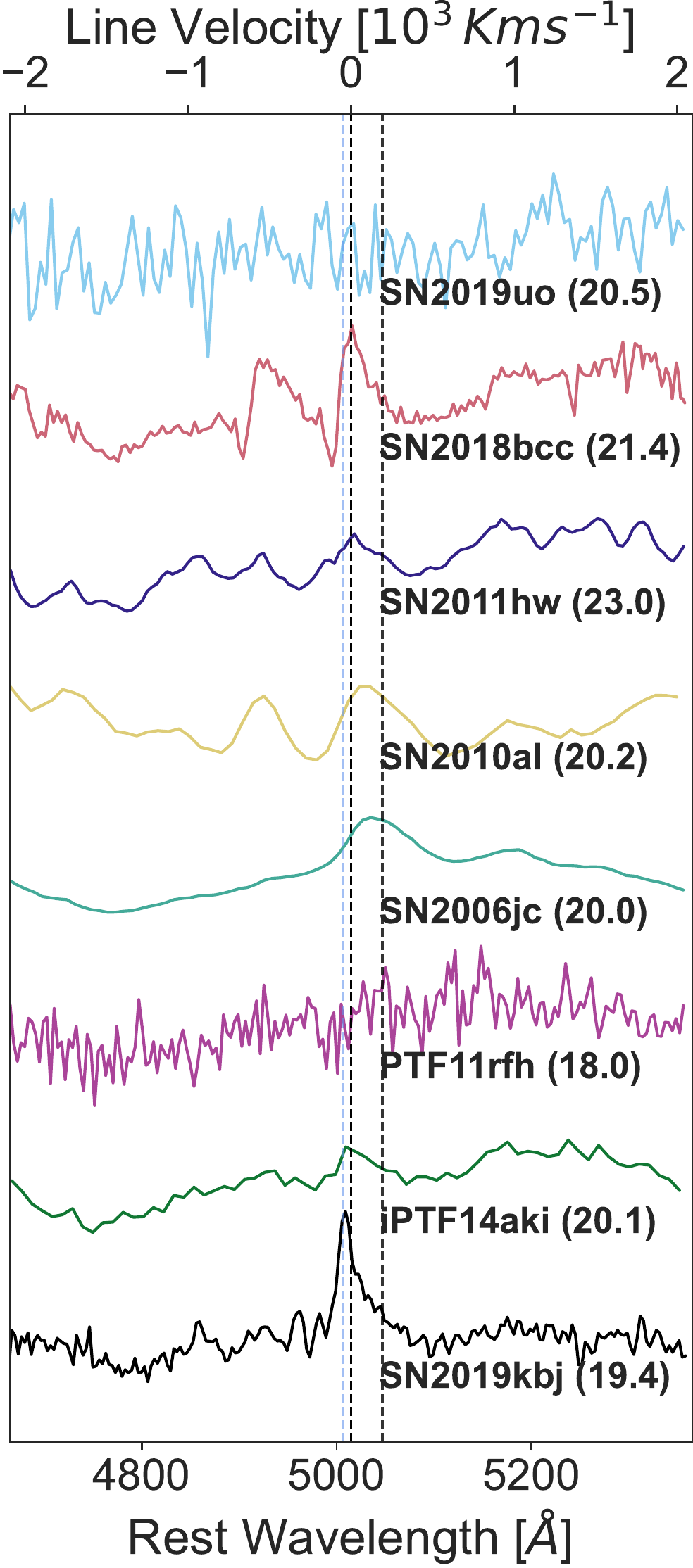} &
    \includegraphics[width=0.2\textwidth]{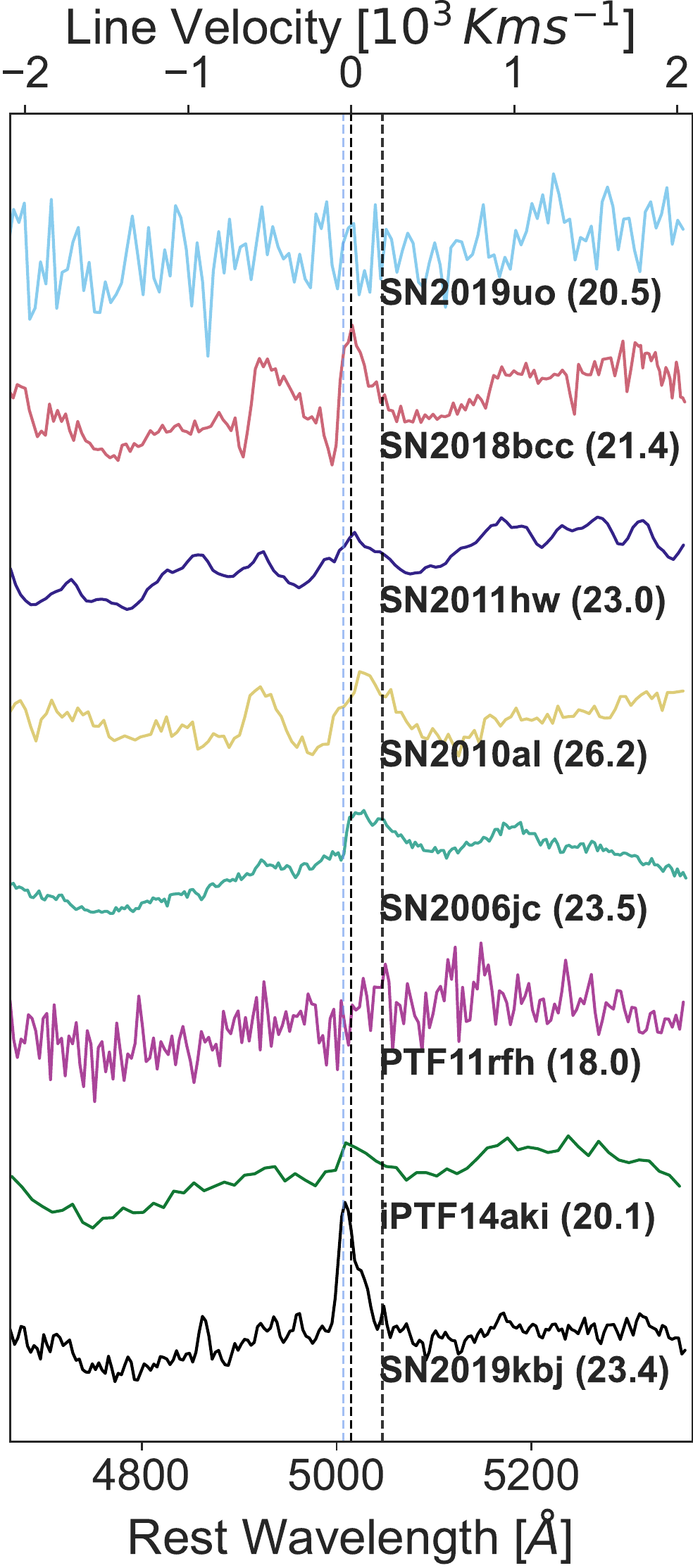}
    \end{tabular}
    
    \caption{The He I $5015\text{\AA}$ and $5047\text{\AA}$ blend (dashed black lines) in SN\,2019kbj, compared to other Type Ibn SNe, at various epochs. The top axis denotes the velocity relative to $5015\text{\AA}$. All spectra are continuum subtracted. For each epoch, only spectra taken \textpm 5 days compared to the phase of the spectra of SN\,2019kbj are compared. The light blue dashed line denotes the O III 5007\AA\ line (which we attribute to the host galaxy).}
    \label{fig:spec2}
\end{figure*}

The spectroscopic evolution of SN\,2019kbj is shown in the left panel of Figure \ref{fig:spec1}. A blue continuum is seen in the early spectra, with no prominent flash-spectroscopy features. However, our earliest spectrum was obtained 2.41 days before peak, which is later than when flash features were observed in other Type Ibn events. In SN\,2019uo \citep{Gangopadhyay_2020} prominent flash features were seen only up to 3.7 days before peak, while in SN\,2010al \citep{pastorello15a} flash features were seen 8 days before peak and disappeared four days later. Since our earliest spectrum is later than these times, we cannot rule out the existence of flash features for SN\,2019kbj.

\begin{figure}
    \includegraphics[width=0.48\textwidth]{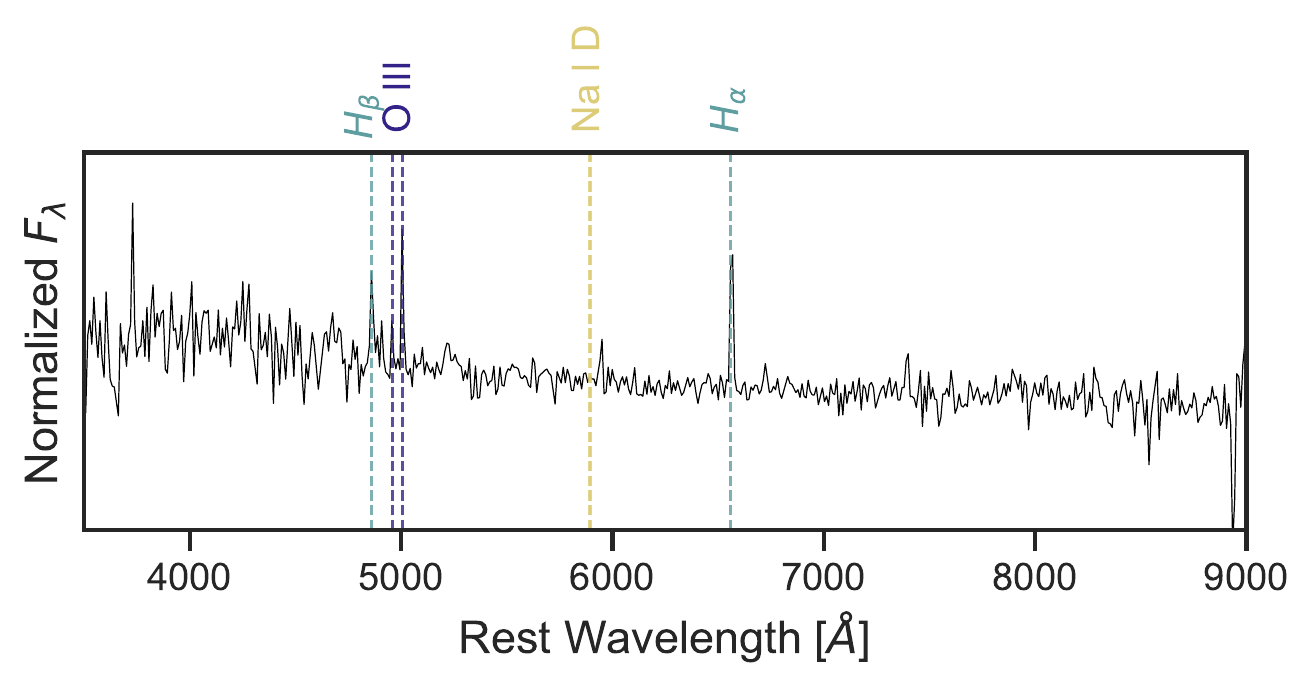}
    \caption{A spectrum of the SN host galaxy taken after the SN faded. Narrow H$\alpha$ and \ion{O}{3}, seen in also in the spectra taken while the SN was active, are present here and thus attributed to the host galaxy. No strong Na I D absorption is detected, indicating little or no host-galaxy extinction.}
    \label{fig:galactic_spec}
\end{figure}

Prominent narrow \ion{He}{1} emission lines can be seen throughout the evolution, together with \ion{Si}{2}, \ion{Mg}{1} and \ion{Ca}{2} lines, which are seen in other Type Ibn SNe as well (right panel of Figure \ref{fig:spec1}). A very prominent \ion{He}{1} blend at $5015$ and $5047\text{\AA}$ develops shortly after peak. This blend is also seen in some other Type Ibn SNe (Fig. \ref{fig:spec2}). It appears after peak, and at later times the $5047\text{\AA}$ component disappears. The $5015\text{\AA}$ component is further blended with O III $5007\text{\AA}$ which we attribute to the underlying host galaxy (Fig. \ref{fig:galactic_spec}).

\begin{figure}
    \includegraphics[width=0.48\textwidth]{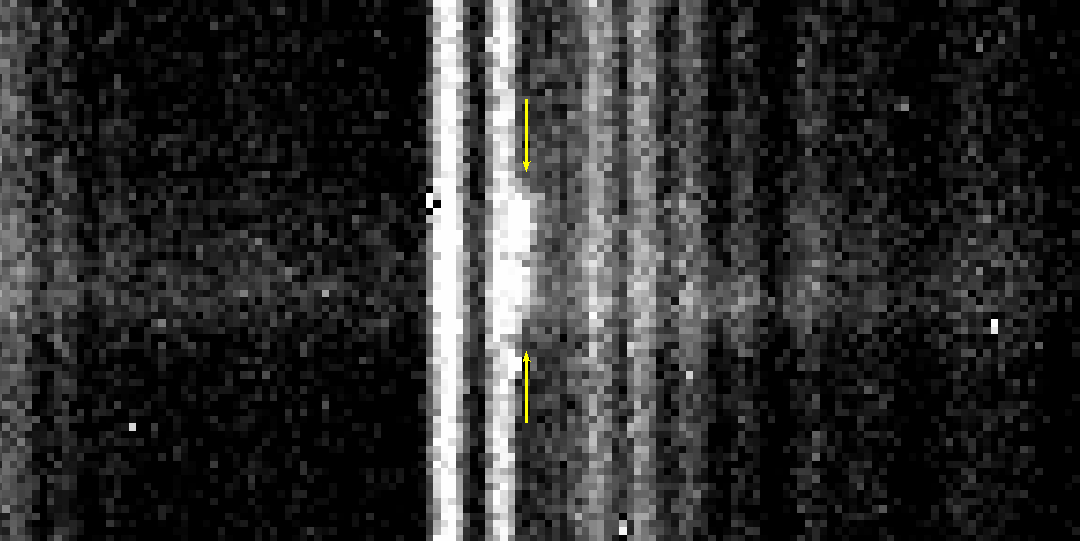}
    \caption{The H$\alpha$ region of the two-dimensional calibrated spectrum of SN\,2019kbj taken on 2019 August 09, with the wavelength axis along the horizontal direction and the spatial axis along the vertical direction. The H$\alpha$ line, marked with yellow arrows, is adjacent to a sky line but can be clearly seen as an extended emission feature. We thus attribute it to the host galaxy.}
    \label{fig:spec_2d}
\end{figure}

Narrow H$\alpha$ emission is also seen in all epochs. This feature may be from the host galaxy or from H in the CSM surrounding the SN progenitor. Here, we attribute the narrow H$\alpha$ emission to the host galaxy for a few reasons. First, it becomes stronger relative to other features as the SN fades. Additionally, it is seen in our host-galaxy spectrum (Fig. \ref{fig:galactic_spec}) and as an extended feature in the two-dimensional spectra of our SNe (an example two-dimensional spectrum is shown in Figure \ref{fig:spec_2d}).
Since the host and SN spectra were each taken under different seeing conditions and with different slit orientations, it is not possible to accurately isolate the amount of H$\alpha$ or O III emission contributed by the host galaxy to each SN spectrum. Therefore, we can neither robustly associate nor rule out an association of a small amount of H$\alpha$ or O III with the SN.

\begin{figure}[hbt]
    \includegraphics[width=0.48\textwidth]{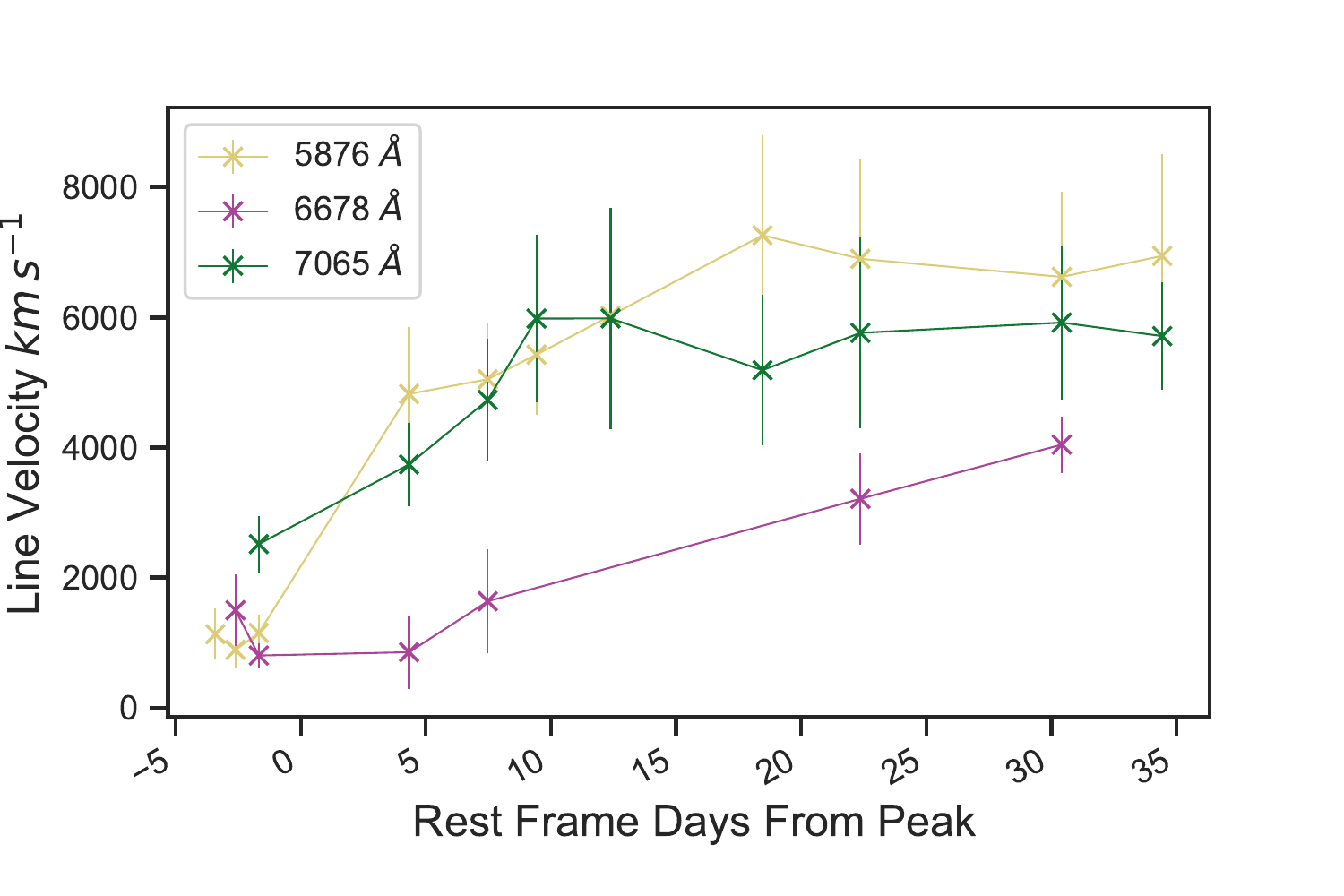}
    \caption{He I expansion velocities measured from the P-cygni minima of each line. These velocities and their time evolution are similar to those seen in other Type Ibn SNe \citep[][and references therein]{Gangopadhyay_2020}.}
    \label{fig:vel_ev}
\end{figure}

We measure expansion velocites from the He I $5876$, $6678$, and $7065 \textrm{\AA}$ lines as was done for SN\,2019uo by \cite{Gangopadhyay_2020}. We first normalize the spectra with a parabolic fit to the continuum and then model each He P Cygni absorption line with a Gaussian. The offset between the best-fit Gaussian center and the line rest-frame wavelength is then translated to an expansion velocity. Our results are shown in Figure \ref{fig:vel_ev}. All lines show expansion velocities of a few $10^3$\,km\,s$^{-1}$ which increase with time during the first $\sim$30 days after peak. This is the same behavior seen in the sample of Ibn SNe analyzed by \cite{Gangopadhyay_2020} and references therein.

\section{Light-curve Modeling} \label{sec:bol_m}

\subsection{Blackbody Fits and Bolometric Light Curve}

\begin{figure}
    \includegraphics[width=0.5\textwidth]{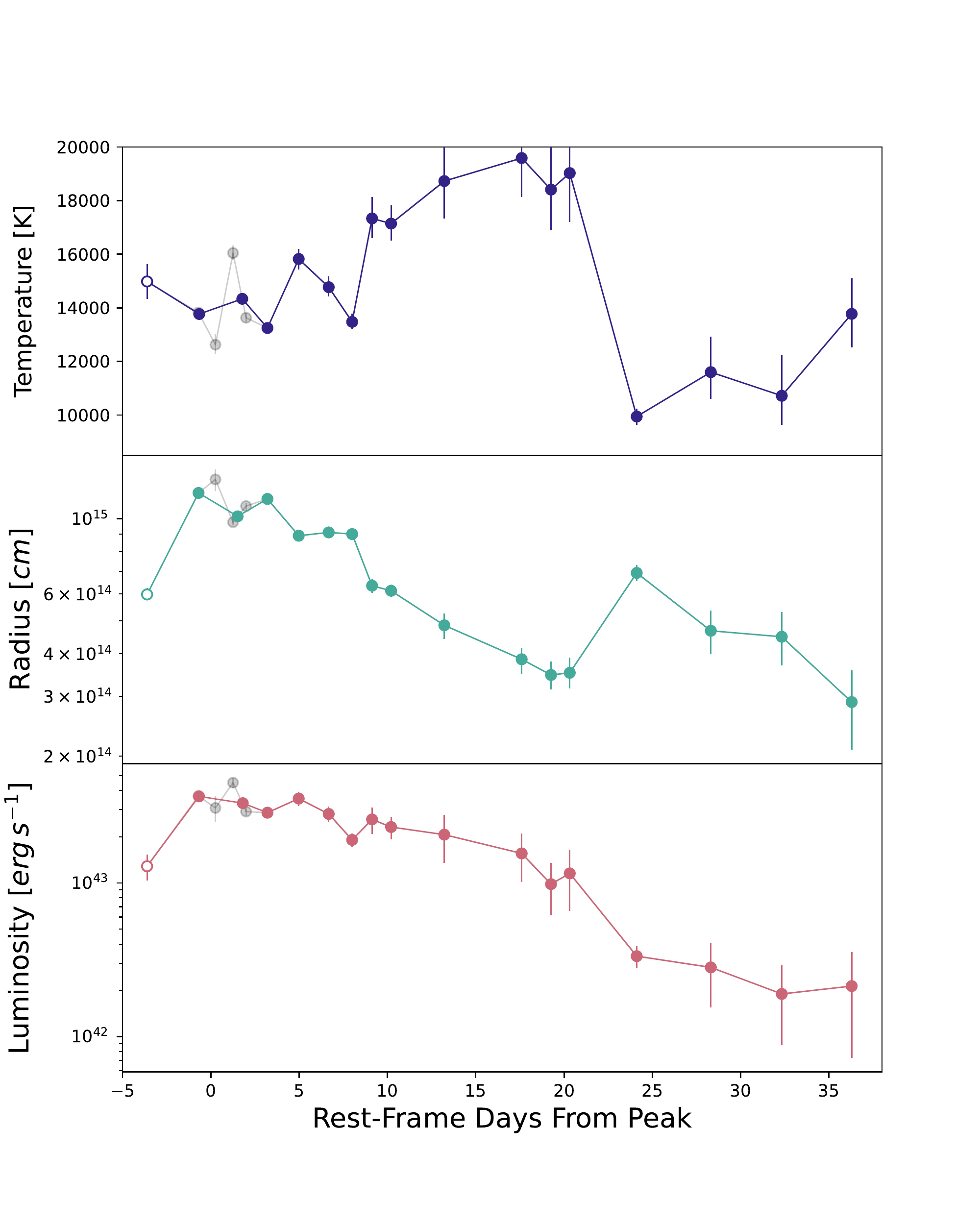}
    \caption{The blackbody temperature, radius, and inferred bolometric luminosity of SN\,2019kbj. For the first epoch (empty symbol), where only one band is available, we assume the temperature to be equal to the average temperature during the rest of the evolution, and use our single-band data at that epoch to constrain the radius and hence bolometric luminosity there. The unbinned data are shown in black semitransparent points.}
    \label{fig:bolphot}
\end{figure}

We fit the spectral energy distribution of each epoch to a blackbody using a Markov Chain Monte Carlo (MCMC) routine implemented via the \texttt{lightcurve\_fitting}\footnote{\url{https://github.com/griffin-h/lightcurve_fitting}} Python library \citep{lc_fit}. We include only epochs with at least three distinct bands observed within one day and bin data taken within less than one day of each other. The best-fit blackbody temperature and radius for each epoch, together with the resulting bolometric luminosity, are presented in Figure \ref{fig:bolphot} and Table \ref{tab:boldat}. As expected from the roughly constant colors, the temperature is seen to be roughly constant at $\sim$15,000\,K out to about a month after peak luminosity. These temperatures are low enough that we do not have to limit ourselves to epochs with ultraviolet coverage to ensure we are correctly sampling the blackbody spectrum \citep{Arcavi2022blackbodies}. 

We add a bolometric epoch prior to peak where we only have the single $o$-band discovery detection, assuming the measured constant temperature can be extrapolated backward to that epoch. We assume a temperature of $14,984\pm650$\,K (the average temperature and its standard deviation from all subsequent epochs)  to calculate a bolometric correction for the $o$-band data point using the \texttt{synphot}\footnote{\url{https://github.com/spacetelescope/synphot_refactor.git}} package \citep{synphot}. This epoch is presented with an empty symbol in Figures \ref{fig:bolphot}--\ref{fig:CSMdist}.

\begin{deluxetable}{hcccc}
\tablecaption{Results of blackbody fits to the photometry of SN\,2019kbj.} \label{tab:boldat}
\tablehead{
\nocolhead{MJD} & \colhead{Phase} & \colhead{Temperature} & \colhead{Radius} & \colhead{Luminosity} \\
\nocolhead{} & \colhead{($days$)} & \colhead{($K$)} & \colhead{($10^{14} cm$)} & \colhead{($10^{43} erg\,s^{-1}$)}
}
\startdata
58665.5 & -3.61 & $14984_{-650}^{+650}$ & $5.98_{-0.24}^{+0.26}$ & $1.285_{-0.245}^{+0.250}$\\
58668.4 & -0.7 & $13824_{-80}^{+89}$ & $11.89_{-0.09}^{+0.09}$ & $3.681_{-0.103}^{+0.110}$\\
58669.4 & 0.26 & $12626_{-355}^{+411}$ & $13.04_{-0.98}^{+0.93}$ & $3.080_{-0.579}^{+0.594}$\\
58670.4 & 1.26 & $16047_{-274}^{+282}$ & $9.76_{-0.19}^{+0.19}$ & $4.499_{-0.353}^{+0.361}$\\
58671.1 & 2.0 & $13630_{-178}^{+179}$ & $10.89_{-0.26}^{+0.25}$ & $2.915_{-0.205}^{+0.205}$\\
58672.3 & 3.21 & $13248_{-135}^{+138}$ & $11.43_{-0.17}^{+0.18}$ & $2.867_{-0.145}^{+0.148}$\\
58674.1 & 4.98 & $15823_{-384}^{+369}$ & $8.91_{-0.22}^{+0.25}$ & $3.542_{-0.386}^{+0.385}$\\
58675.8 & 6.68 & $14771_{-353}^{+408}$ & $9.11_{-0.29}^{+0.28}$ & $2.813_{-0.322}^{+0.354}$\\
58677.1 & 8.0 & $13478_{-271}^{+310}$ & $9.01_{-0.26}^{+0.24}$ & $1.907_{-0.188}^{+0.203}$\\
58678.2 & 9.13 & $17336_{-737}^{+803}$ & $6.35_{-0.30}^{+0.31}$ & $2.591_{-0.505}^{+0.542}$\\
58679.3 & 10.21 & $17143_{-634}^{+689}$ & $6.13_{-0.25}^{+0.26}$ & $2.313_{-0.391}^{+0.420}$\\
58682.3 & 13.22 & $18728_{-1388}^{+1828}$ & $4.85_{-0.42}^{+0.40}$ & $2.063_{-0.710}^{+0.875}$\\
58686.7 & 17.6 & $19587_{-1451}^{+2083}$ & $3.85_{-0.36}^{+0.32}$ & $1.557_{-0.544}^{+0.710}$\\
58688.4 & 19.26 & $18411_{-1496}^{+1876}$ & $3.46_{-0.32}^{+0.33}$ & $0.983_{-0.367}^{+0.441}$\\
58689.4 & 20.32 & $19027_{-1822}^{+2119}$ & $3.52_{-0.35}^{+0.39}$ & $1.154_{-0.499}^{+0.574}$\\
58693.2 & 24.11 & $9946_{-304}^{+287}$ & $6.92_{-0.37}^{+0.38}$ & $0.334_{-0.054}^{+0.053}$\\
58697.4 & 28.31 & $11602_{-996}^{+1330}$ & $4.67_{-0.68}^{+0.70}$ & $0.282_{-0.127}^{+0.154}$\\
58701.4 & 32.33 & $10719_{-1083}^{+1517}$ & $4.49_{-0.79}^{+0.82}$ & $0.189_{-0.102}^{+0.128}$\\
58705.4 & 36.28 & $13775_{-3247}^{+5326}$ & $2.88_{-1.00}^{+1.32}$ & $0.213_{-0.250}^{+0.383}$\\
\enddata
\tablecomments{For the first epoch, where only one band is available, we assume the temperature to be equal to the average temperature during the rest of the evolution, and use our single-band data at that epoch to constrain the radius and hence bolometric luminosity there.}
\end{deluxetable}

\subsection{Modeling the Bolometric Light Curve}

We fit the bolometric light curve of SN\,2019kbj to two models: the radioactive decay model from \cite{arnett82}, \cite{valenti08}, and \cite{chatzopoulos12}, and the radioactive decay model with additional CSM-interaction power from \cite{chevalier82} and \cite{chatzopoulos12}. Each model is fit to the data using the \texttt{Bolometric\_Modelling}\footnote{\url{https://github.com/Tomariebenami/Bolometric_Modelling}} module \citep{bolmod}.

\subsubsection{Radioactive Decay Model}\label{sec:rd}
The radioactive decay model assumes that the bolometric luminosity is powered solely by the radioactive decay of $^{56}$Ni to $^{56}$Co to $^{56}$Fe with $\gamma$-ray leakage taken into consideration \citep{valenti08, chatzopoulos12}. The luminosity is given by
\begin{eqnarray}
L(t) & = & \frac{M_{Ni}}{t_{m}} e^{\frac{t}{t_{m}}} \nonumber \\
     & \times & \left[\left(\epsilon_{Co} - \epsilon_{Ni}\right)\int_{0}^{x} B(z) \,dz + \epsilon_{Co}\int_{0}^{x} C(z)\,dz \right] \nonumber \\
     & \times & \left(1 - e^{At^{-2}}\right) 
\end{eqnarray}
where the free parameters are the ejecta mass $M_{ej}$, the $^{56}$Ni mass $M_{Ni}$, the characteristic ejecta velocity $v_{ej}$, and the optical opacity $\kappa_{opt}$ (we also fit for the explosion time relative to the peak, $t_0$). $\epsilon_{Ni} = 3.90\times 10^9$ erg\,s$^{-1}$\,g$^{-1}$ and $\epsilon_{Co} = 6.78\times 10^9$ erg\,s$^{-1}$\,g$^{-1}$ are the energy-generation rates of the decays of $^{56}$Ni and $^{56}$Co respectively \citep{sutherland84, cappellaro97}. The integrands $B(z) = 2z e^{-2zy + z^2}$ and $C(z) = 2z e^{-2zy + 2zs + z^2}$ are the luminosity outputs of the decays of $^{56}$Ni and $^{56}$Co, with $y = t_m / 2\tau_{Ni}$ and $s = t_m(\tau_{Co} - \tau_{Ni})/(2\tau_{Co}\tau_{Ni})$. Both integrals are evaluated up to $x = t / t_m$ with
\begin{equation}
t_m = \left(\frac{\kappa_{opt}}{\beta c}\right)^{1/2} \left( \frac{20 M_{ej}^2}{3 v_{ej}^2}\right)^{1/4}
\end{equation}
defined as the light curve timescale, and $\tau_{Ni} = 8.8$ days and $\tau_{Co} = 111.3$ days the respective decay lifetimes \citep[e.g.][]{decayprop}.  Finally, 
\begin{equation}
A = \frac{3 \kappa_{\gamma} M_{ej}}{4\pi v_{ej}^2}
\end{equation}
is the $\gamma$-ray leakage factor. We set the $\gamma$-ray opacity, $\kappa_{\gamma}$, to $0.027$\,cm$^{2}$g$^{-1}$ following \cite{swartz95} and \cite{cappellaro97}. 

We use the MCMC fitting method, implemented through the \texttt{emcee} package \citep{emcee}, with 500 burn-in steps, followed by 8,000 fitting steps with 150 walkers. We limit the $^{56}$Ni mass to be less than the total ejecta mass and use a very broad ejecta velocity ($v_{ej}$) prior, since otherwise the fit prefers an unphysical solution with more $^{56}$Ni than total ejecta mass (see Table \ref{tab:uc_RD_p} in Appendix \ref{apend:B1}; a similar result was obtained for the extremely luminous Ibn SN ASASSN-14ms by \citealt{Vallely2018} and \citealt{wang21}). 

\begin{figure}
    \includegraphics[width=0.5\textwidth]{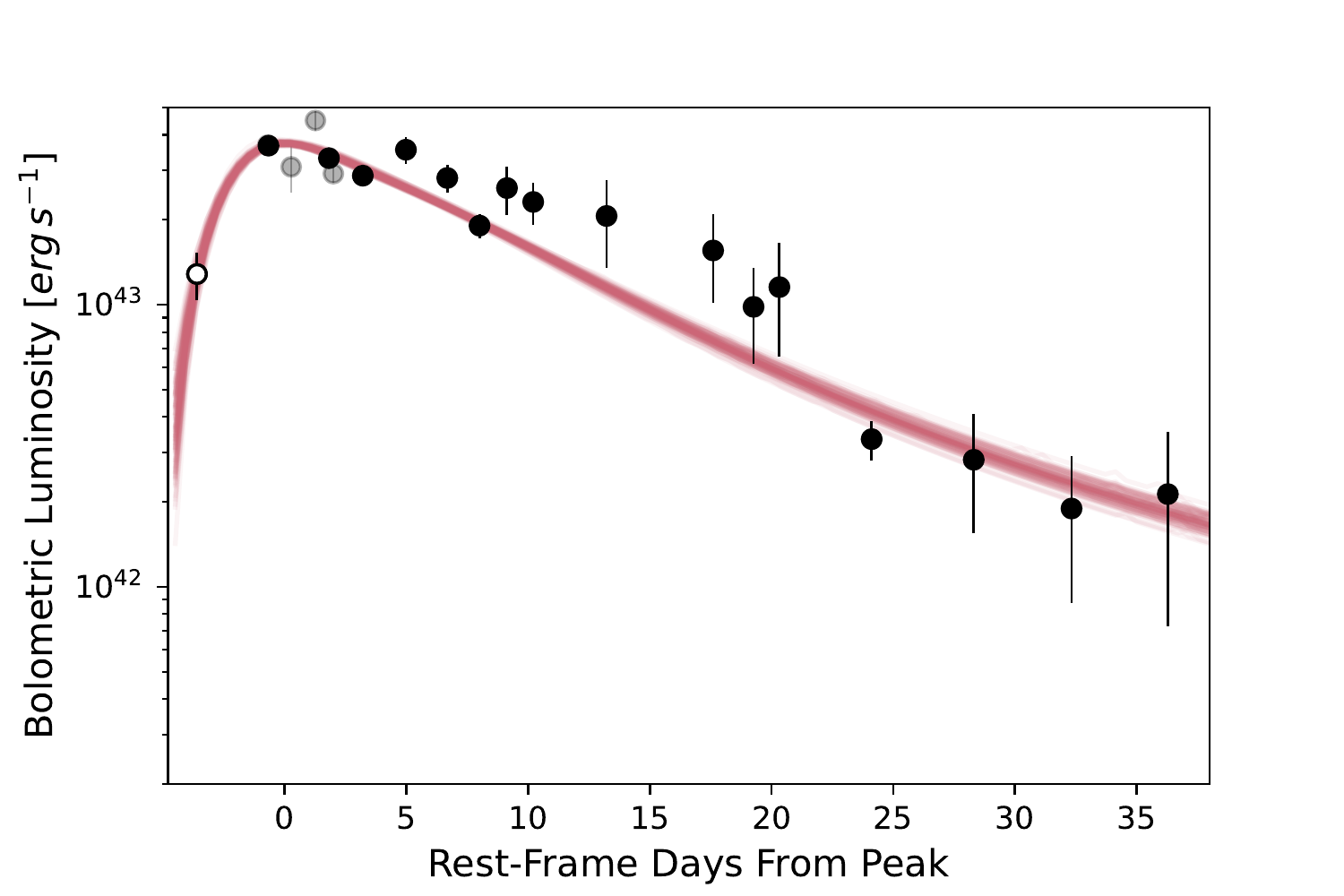}
    \caption{Radioactive decay model (100 lines, chosen at random from the MCMC walker distribution) compared to the bolometric light curve of SN\,2019kbj (binned data in opaque points, unbinned data in semitransparent points).}
    \label{fig:RD}
\end{figure}

Our fit is shown in Figure \ref{fig:RD} and the best-fit parameters are given in Table \ref{tab:RD_p}. The corner plot of the fit is shown in Figure \ref{fig:RD_corner} in Appendix \ref{apend:B}. Although we find a reasonable fit to the data, it requires a very high ejecta velocity (of order 60,000\,km\,s$^{-1}$), which is not typically seen in any type of SN. In addition, it requires a large $^{56}$Ni mass of $\sim$0.8\,$M_{\odot}$, which is also not typical of core-collapse SNe. The ejecta mass remains unconstrained within the prior bounds. We conclude that radioactive decay is disfavored as the sole power source of the light curve of SN\,2019kbj. 

Given the long-lived blue continuum and narrow He lines in Type Ibn SNe, CSM interaction is a most likely additional source of power.

\begin{deluxetable}{ccccc}[hb]
\tabletypesize{\footnotesize}
\tablecaption{Best-fit parameters for the radioactive decay model.} \label{tab:RD_p}
\tablehead{
\colhead{$M_{Ni}$} & \colhead{$M_{ej}$} & \colhead{$v_{ej}$} & \colhead{$\kappa_{opt}$} & \colhead{$t_{0}$}\\
\colhead{($M_{\odot}$)} & \colhead{($M_{\odot}$)} & \colhead{($10^3 km\,s^{-1}$)} & \colhead{($cm^2 g^{-1}$)} & \colhead{($days$)}}
\startdata
$0.776_{-0.015}^{0.016}$ & $10.08_{-6.60}^{6.75}$ & $59.05_{-24.34}^{17.20}$ & $0.0040_{-0.0010}^{0.0028}$ & $-5.32_{-0.17}^{0.16}$ \\
\enddata
\end{deluxetable}

\subsubsection{Radioactive Decay + Circumstellar Material Interaction Model}

We next fit a radioactive decay model with additional CSM-interaction power as formulated by \cite{chevalier82} and \cite{chatzopoulos12}. In this model the CSM density, $\rho_{csm}$, is described by a power law, $\rho_{csm} = qr^{-s}$. The ejecta distribution is described by two power laws, $\rho_{ej} \propto r^{-\delta}$ for $r$ smaller than a critical normalized radius $x_0$, and $\rho_{ej} \propto r^{-n}$ for $r$ larger than $x_0$. The total SN luminosity in this model is given by \citep{chatzopoulos12}

\begin{equation} \label{eq:rd+cdm}
	\begin{split}
	L(t) & = \frac{1}{t_{0}}e^{-\frac{t}{t_{0}}}\int_{0}^{t}e^{\frac{t'}{t_{0}}} [ \frac{2\pi}{(n-s)^3)} g^{n\frac{5-s}{n-s}} q^\frac{n-5}{n-s} (n-3)^2(n-5) \\
	 & \times \beta_F^{(5-s)} A^{\frac{5-s}{n-s}}(t'-t_i)^{\frac{2n+6s-ns-15}{n-s}}\theta(t_{FS, BO} - t') \\
	 & + 2\pi \left ( \frac{Ag^n}{q} \right )^{\frac{5-n}{n-s}} \beta_R^{5-n}g^n \left ( \frac{3-s}{n-s} \right ) \\
	 & \times (t' + t_i)^{\frac{2n+6s-ns-15}{n-s}} \theta(t_{RS,*} - t') ] \\
	 & + \frac{1}{t'_{0}}e^{-\frac{t}{t'_{0}}} \\
	 & \times \int_{0}^{t} e^{\frac{t'}{t'_0}}M_{Ni}\left [ (\epsilon_{Ni} - \epsilon_{Co})e^{-\frac{t'}{t_{Ni}}} + \epsilon_{Co} e^{-\frac{t'}{t_{Co}}} \right ] dt'
	\end{split}
\end{equation}

The model has 10 free parameters (in addition to the density power-law indices $s$, $\delta$ and $n$, which we fix): the ejecta mass $M_{ej}$, the $^{56}$Ni mass $M_{Ni}$, the CSM mass $M_{CSM}$, the characteristic ejecta velocity $v_{ej}$, the density of the CSM shell, $\rho_{CSM, in}$, at its innermost radius $r_{in}$ ($q=\rho_{CSM, in}r_{in}^s$), the efficiency of converting shock energy to luminosity $\epsilon$, the normalized radius at which the ejecta power law switches indices $x_{0}$, the optical opacity $\kappa_{opt}$, and the explosion time $t_{0}$, relative to peak time. 

In addition to the free parameters, $\theta(t_{FS} -t)$ and $\theta(t_{RS} -t)$ are Heaviside step functions corresponding to the termination of the forward and reverse shock waves at times $t_{FS}$ and $t_{RS}$, respectively, which are dictated by the free parameters (see \citealt{chatzopoulos12} and \citealt{chevalier82} for the full details), $g(n, \delta, M_{ej}, v_{ej})$ is a scaling parameter for the ejecta density, and $\beta_R$,  $\beta_F$ and $A$ are all constants found in \cite{chevalier82}. All parameters related to the radioactive decay component (the second integral in Equation \ref{eq:rd+cdm}) are identical to those of the radioactive decay model described in Section \ref{sec:rd}.

Here, we study two cases: $s=0$ (a uniform-density CSM shell) and $s=2$ (CSM due to a steady mass-loss wind). For each case, we test both $\delta=0$ and $\delta=2$, which are values typically assumed for the inner density profile in SNe \citep{chatzopoulos12}. As for the outer density profile, $n$, previous works used $n=10$ and $n=12$ \citep{Gangopadhyay_2020,Pellegrino2022}. However, \cite{chatzopoulos12} mention that $n=11.7$ corresponds to red supergiant progenitors \citep{Matzner1999}, while more compact stars (such as the stripped-envelope progenitors expected for Type Ibn SNe) are characterized by lower values of $n$. Therefore, $n=12$ is probably not appropriate for Type Ibn SN progenitors, and even $n=10$ might be too high. \cite{Chevalier1981,chevalier82} find that $n=7$ is able to reproduce light curves of Type Ia supernovae from white dwarf progenitors. Therefore the true value of $n$ for Type Ibn progenitors is possibly somewhere between 7 and 10. Here we test both edge values. In summary we test all combinations of $n=7,10$, $s=0,2$, and $\delta=0,2$.

Given the large number of parameters, we fit each case using the dynamic nested sampling method as implemented by the \texttt{DYNESTY} Python package \citep{dynesty}. We use uniform and log-uniform priors, as detailed in Appendix \ref{apend:A}. We require the $^{56}$Ni mass to be less than the total ejecta mass in all fits.

\begin{figure*}
    \includegraphics[width=\textwidth]{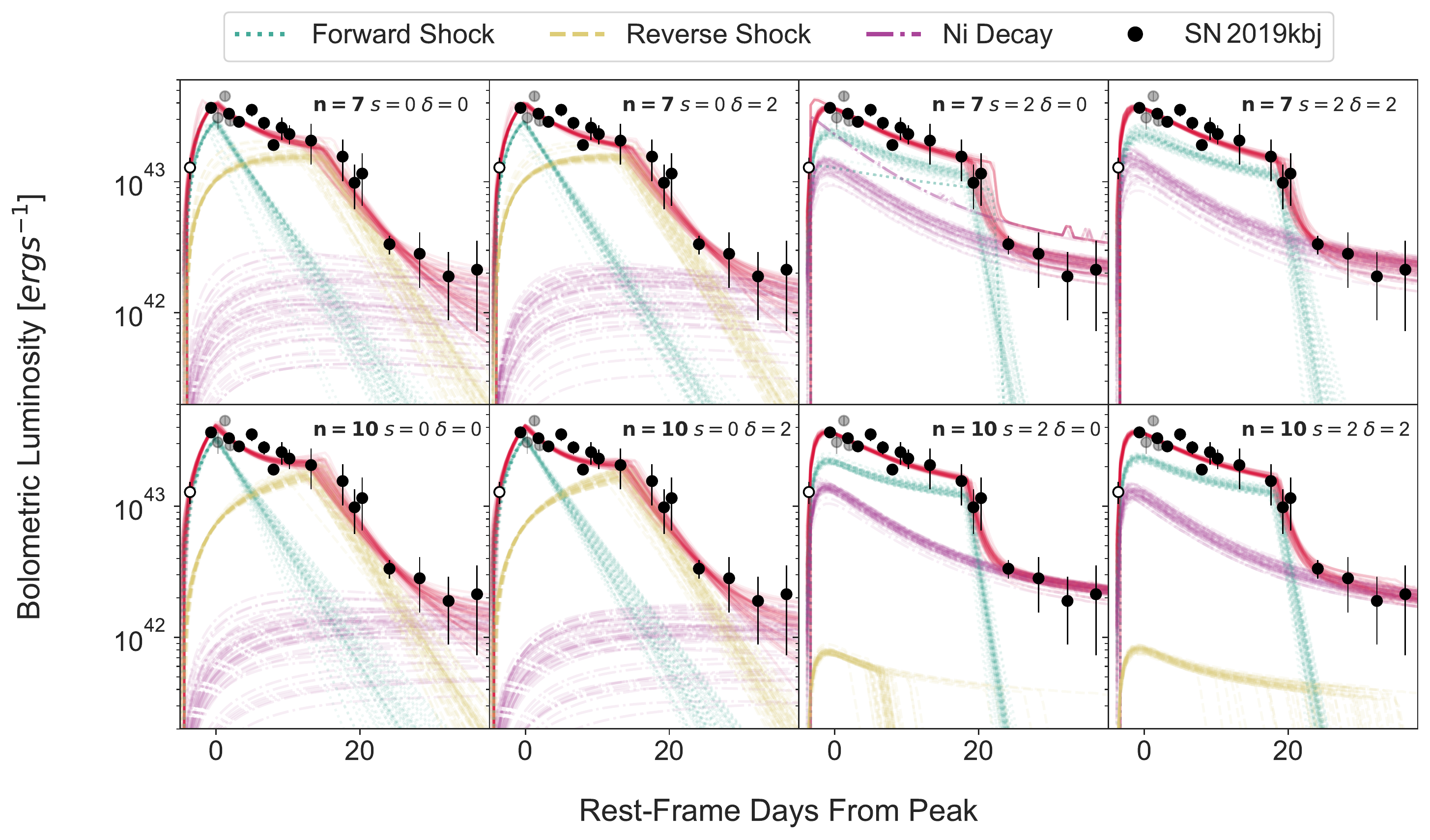}
    \caption{Radioactive-decay + CSM interaction model (50 lines, chosen at random from the sampler's distribution) compared to the bolometric light curve of SN\,2019kbj (binned data in opaque points, un-binned data in semi-transparent points) for different values of the fixed parameters. For the $n=7,\,s=2$ cases, the reverse shock contribution is $\lesssim10^{40} erg/\, s^{-1}$, and is not shown. Small bumps seen in some models at $\sim$30 days are numerical artifacts from the integration.}
    \label{fig:CSMdist}
\end{figure*}

Our fits are shown in Figure \ref{fig:CSMdist}, best-fit parameters given in Table \ref{tab:RDCSM_p}, and corner plots are shown in Figures \ref{fig:cp_n7d0s0}--\ref{fig:cp_n10d2s2} in Appendix \ref{apend:B}.

\begin{deluxetable*}{ccccccccccccc}
\tablecaption{Best-fit parameters for the radioactive decay + CSM interaction model.} \label{tab:RDCSM_p}
\tablewidth{0pt}
\tablehead{
\colhead{$n$} & \colhead{$s$} & \colhead{$\delta$} & \colhead{$M_{ej}$} & \colhead{$M_{Ni}$} & \colhead{$M_{csm}$} & \colhead{$v_{ej}$} & \colhead{$\rho_{csm, in}$} & \colhead{$r_{in}$} & \colhead{$\epsilon$} & \colhead{$x_{0}$} & \colhead{$\kappa_{opt}$} & \colhead{$t_{0}$}\\
& & & \colhead{($M_{\odot}$)} & \colhead{($M_{\odot}$)} & \colhead{($M_{\odot}$)} & \colhead{($10^3 km\,s^{-1}$)} & \colhead{($10^{-12}g\,cm^{-3}$)} & \colhead{($10^{14}\,cm$)} &  &  & \colhead{($cm^2 g^{-1}$)} & \colhead{($days$)}} 
\startdata
    7 & 0 & 0 & $0.76_{-0.38}^{+0.42}$ & $0.08_{-0.03}^{+0.03}$ & $0.13_{-0.06}^{+0.07}$ & $9.34_{-2.22}^{+5.33}$ & $0.93_{-0.65}^{+1.99}$ & $9.20_{-5.23}^{+8.99}$ & $0.56_{-0.24}^{+0.23}$ & $0.60_{-0.17}^{+0.27}$ & $0.79_{-0.23}^{+0.15}$ & $-4.62_{-0.17}^{+0.19}$ \\
     & & 2 & $0.78_{-0.28}^{+0.39}$ & $0.08_{-0.03}^{+0.02}$ & $0.13_{-0.04}^{+0.06}$ & $14.58_{-4.33}^{+3.01}$ & $0.73_{-0.55}^{+1.81}$ & $10.59_{-6.10}^{+5.91}$ & $0.46_{-0.28}^{+0.36}$ & $0.57_{-0.17}^{+0.24}$ & $0.78_{-0.26}^{+0.15}$ & $-4.63_{-0.16}^{+0.15}$ \\
    \hline 
     & 2 & 0 & $0.42_{-0.10}^{+0.09}$ & $0.23_{-0.03}^{+0.03}$ & $0.29_{-0.13}^{+0.12}$ & $14.02_{-2.15}^{+1.80}$ & $0.11_{-0.08}^{+0.10}$ & $5.17_{-1.41}^{+3.62}$ & $0.19_{-0.06}^{+0.10}$ & $0.22_{-0.05}^{+0.06}$ & $0.21_{-0.07}^{+0.15}$ & $-3.97_{-0.07}^{+0.12}$ \\
     & & 2 & $0.28_{-0.05}^{+0.11}$ & $0.22_{-0.04}^{+0.04}$ & $0.12_{-0.02}^{+0.05}$ & $13.00_{-4.08}^{+4.95}$ & $0.13_{-0.11}^{+0.19}$ & $3.17_{-1.35}^{+5.21}$ & $0.41_{-0.14}^{+0.16}$ & $0.25_{-0.07}^{+0.19}$ & $0.66_{-0.29}^{+0.21}$ & $-3.97_{-0.10}^{+0.14}$ \\ 
    \hline
    10 & 0 & 0 &  $1.19_{-0.31}^{+0.57}$ & $0.10_{-0.04}^{+0.05}$ & $0.06_{-0.02}^{+0.03}$ & $11.25_{-3.18}^{+4.32}$ & $0.76_{-0.49}^{+1.45}$ & $3.92_{-2.53}^{+3.20}$ & $0.61_{-0.25}^{+0.26}$ & $0.54_{-0.22}^{+0.21}$ & $0.76_{-0.19}^{+0.17}$ & $-4.95_{-0.35}^{+0.22}$ \\
     & & 2 & $1.42_{-0.52}^{+0.94}$ & $0.10_{-0.03}^{+0.04}$ & $0.07_{-0.02}^{+0.04}$ & $11.87_{-3.84}^{+4.18}$ & $0.65_{-0.39}^{+1.75}$ & $4.52_{-2.60}^{+4.56}$ & $0.46_{-0.22}^{+0.30}$ & $0.61_{-0.20}^{+0.21}$ & $0.78_{-0.23}^{+0.15}$ & $-4.95_{-0.27}^{+0.22}$ \\
    \hline 
     & 2 & 0 & $0.26_{-0.02}^{+0.03}$ & $0.23_{-0.02}^{+0.01}$ & $0.21_{-0.03}^{+0.03}$ & $13.53_{-0.94}^{+2.80}$ & $1.20_{-0.98}^{+0.53}$ & $1.72_{-0.21}^{+1.27}$ & $0.61_{-0.15}^{+0.08}$ & $0.45_{-0.04}^{+0.12}$ & $0.14_{-0.03}^{+0.13}$ & $-3.95_{-0.05}^{+0.05}$ \\
     & & 2 & $0.23_{-0.02}^{+0.04}$ & $0.21_{-0.02}^{+0.02}$ & $0.08_{-0.01}^{+0.02}$ & $16.05_{-3.32}^{+0.95}$ & $0.60_{-0.36}^{+0.29}$ & $1.32_{-0.15}^{+0.65}$ & $0.96_{-0.08}^{+0.03}$ & $0.48_{-0.04}^{+0.19}$ & $0.54_{-0.13}^{+0.13}$ & $-3.92_{-0.06}^{+0.06}$ \\
\enddata
\end{deluxetable*}

\section{Discussion} \label{sec:disc}

SN\,2019kbj is photometrically and spectroscopically similar to other Type Ibn SNe, with a post-peak luminosity decline rate that is similar to that of the average Type Ibn light curve (Fig. \ref{fig:phot2}).

The blackbody temperature of SN\,2019kbj is relatively constant around $15,000$\,K, in contrast to the cooling seen in H-rich Type II SNe, for example \citep[e.g.][]{Valenti2014}. This indicates that an ongoing power source, likely CSM interaction, continues to heat the ejecta as it expands. A constant color, indicative of constant temperature, is seen also in other Type Ibn's (but not all; Fig. \ref{fig:phot3}). The constant color and temperature might be an indication of a common CSM-interaction power source for this class of events. Determining why some events do not show constant color requires additional modeling of those data, which we leave to future work.

The bolometric light curve of SN\,2019kbj yields extreme parameter values when fit by radioactive decay alone, but can be fit with much more reasonable values with the addition of CSM-interaction luminosity. 

The steady-wind CSM ($s=2$) models require higher Ni masses and lower ejecta masses compared to a uniform-density CSM shell ($s=0$; Table \ref{tab:RDCSM_p}). In fact, the steady-wind CSM models require most ($\sim$50--90\%) of the ejecta to be Ni, while in the uniform-density CSM model less than 10\% of the ejecta mass is Ni, as seen in most core-collapse SNe. This is the case regardless of the value of $n$ chosen.
Therefore, we conclude that our models show a slight preference for a uniform-density CSM shell over a steady-wind CSM. This is consistent with the results of \cite{Karamehmetoglu2017} and \cite{Gangopadhyay_2020} who also prefer a uniform-density CSM shell over a steady-wind CSM to explain the light curves of the Type Ibn SNe OGLE-2014-SN-131 and 2019uo.

In the uniform-density CSM case, the bolometric light curve of SN\,2019kbj requires a $^{56}$Ni mass of $0.08$--$0.1M_{\odot}$, which is an order of magnitude higher than the $0.01\,M_{\odot}$ derived by \cite{Gangopadhyay_2020} and \cite{Pellegrino2022} for SN\,2019uo (for both CSM cases). In the steady-wind CSM case, we derive an even higher value of $0.22\,M_{\odot}$ for the $^{56}$Ni mass. The ejecta masses of $\sim0.2$--$1.4\,M_{\odot}$ that we find are substantially lower than the $\sim16\,M_{\odot}$ derived for SN\,2019uo by \cite{Gangopadhyay_2020}, but overlap with the $\sim1\,M_{\odot}$ found by \cite{Pellegrino2022} for that event. 

The ejecta masses deduced are not highly sensitive to the value of $n$, and change only within a factor of two for $n=7$ vs. $n=10$. The $^{56}$Ni and CSM masses are even less sensitive to $n$, and are in fact consistent within the errors for the different $n$ values tested.

The mass-loss rate, $\dot{M}$, that produced the CSM can be obtained from the continuity equation:
\begin{equation}
\begin{split}
    \dot{M}\left(r\right) & = 4{\pi}r^2\rho_{CSM}\left(r\right)v_{w}\left(r\right) \\
            & = 4{\pi}r^2qr^{-s}v_{w}\left(r\right) \\
\end{split}
\end{equation}
where $v_{w}$ is the CSM wind velocity, and we recall that $q=\rho_{CSM,in}r_{in}^s$. If we assume a constant wind velocity, then the mass-loss rate is
\begin{equation}
   \dot{M}\left(r\right) = 4{\pi}r^{2-s}\rho_{CSM,in}r_{in}^sv_{w}
\end{equation}

For the $s=2$ case, this results in a constant $\dot{M}$. Typical Wolf-Rayet wind velocities are of order 1000 km\,s$^{-1}$ \citep{crowther07}, which is consistent with the order of magnitude of the earliest He velocity we measure in Section \ref{sec:spec} (before the CSM might have accelerated significantly due to impact from the SN ejecta). With our $s=2$ best-fit parameters we thus find a mass-loss rate on the order of $\sim$0.5$\left(\frac{v_w}{1000\,\textrm{km\,s}^{-1}}\right)$\,$M_{\odot}$\,yr$^{-1}$ (for both $n=7$ and $n=10$). At this mass-loss rate, to reach the total CSM mass from each fit, the mass-loss episode that shaped the light curve of SN\,2019kbj would have lasted only $\sim$0.2--1.1 years.

The mass-loss rate found here is similar to the mass-loss rate derived for the Type IIn SN iPTF13z (0.1--2\,${M}_{\odot}$\,yr$^{-1}$; \citealt{Nyholm2017}). \cite{Gangopadhyay_2020} found a much higher value for SN\,2019uo ($\sim$200$\left(\frac{v_w}{1000\,\textrm{km\,s}^{-1}}\right)$\,$M_{\odot}$\,yr$^{-1}$) which motivated them to rule out the $s=2$ model for that event.

For $s=0$, the mass loss rate is not constant for a constant $v_w$. Using the best fit parameters for this case, the mass loss rate at the inner CSM radius, $\dot{M}\left(r_{in}\right)$, is $\sim$16$\left(\frac{v_w}{1000\,\textrm{km\,s}^{-1}}\right)$\,$M_{\odot}$\,yr$^{-1}$ for $n=7$ and $\sim$2.5$\left(\frac{v_w}{1000\,\textrm{km\,s}^{-1}}\right)$\,$M_{\odot}$\,yr$^{-1}$ for $n=10$. We can not derive a timescale for the ejection of this shell, without knowing the difference between $r_{in}$ and the progenitor radius (which we can not constrain here). In addition to these uncertainties, $v_w$ might also not be constant in this type of CSM.

Although the light curve of SN\,2019kbj is $\sim$1--2 magnitudes brighter than that of SN\,2019uo, both are within the observed spread of Type Ibn SN luminosities (Fig. \ref{fig:phot2}). Therefore, our results indicate a possible diversity in Type Ibn SN progenitor systems and explosions. However, this apparent diversity may be due, at least in part, to the different model implementation and fitting methods used for each event. As noted previously, \cite{Pellegrino2022} also fit a Ni decay + uniform-density CSM model to SN\,2019uo, but find a very different ejecta mass than that found by \cite{Gangopadhyay_2020}, $\sim1M_{\odot}$ vs. $\sim16M_{\odot}$, using the same data. Comparing physical parameters between events requires fitting their light curves with the same models and methods. We plan to perform such systematic comparisons in future work (T. Ben-Ami et al. 2023, in preparation). 

\section{Summary and Conclusions} \label{sec:conc}

SN\,2019kbj is similar both photometrically and spectroscopically to other Type Ibn SNe. 

We show that the radioactive decay of $^{56}$Ni is likely not enough to explain the light curve, but that an additional power source is needed. This is also evidenced by the roughly constant blackbody temperature of SN\,2019kbj. 

We fit the bolometric light curve of SN\,2019kbj with a radioactive decay + CSM interaction model, and find that a uniform-density CSM shell produces more reasonable fits compared to a steady-wind CSM. 

Other Type Ibn SNe show lower $^{56}$Ni masses and higher ejecta masses compared to SN\,2019kbj. This diversity might be intrinsic to Type Ibn SN progenitor systems and explosions, but it might also arise from differences in the way physical parameters have been inferred for different events. A systematic study of Type Ibn light curves could elucidate this issue and provide additional clues as to the puzzling nature of these events.

~\\
We thank the anonymous referee for useful comments that helped make the results of this paper more robust and insightful. 
T.B.-A acknowledges support from the European Research Council (ERC) under the European Union’s Horizon 2020 research and innovation program (grant No. 852097).
IA is a CIFAR Azrieli Global Scholar in the Gravity and the Extreme Universe program and acknowledges support from that program, from the ERC under the European Union’s Horizon 2020 research and innovation program (grant No. 852097), from the Israel Science Foundation (grant No. 2752/19), from the United States - Israel Binational Science Foundation (BSF), and from the Israeli Council for Higher Education Alon Fellowship. The Las Cumbres group is supported by AST-1911151, AST-1911225, and NASA Swift grant No. 80NSSC19K1639.
This work made use of data from the Las Cumbres Observatory global network of robotic telescopes, and of data from the Asteroid Terrestrial-impact Last Alert System (ATLAS) project. ATLAS is primarily funded to search for near earth asteroids through NASA grant Nos. NN12AR55G, 80NSSC18K0284, and 80NSSC18K1575; by-products of the NEO search include images and catalogs from the survey area. This work was partially funded by Kepler/K2 grant No. J1944/80NSSC19K0112 and HST GO-15889, and STFC grant Nos. ST/T000198/1 and ST/S006109/1. The ATLAS science products have been made possible through the contributions of the University of Hawaii Institute for Astronomy, the Queen’s University Belfast, the Space Telescope Science Institute, the South African Astronomical Observatory, and The Millennium Institute of Astrophysics (MAS), Chile.
This work also made use of the NASA/IPAC Extragalactic Database (NED), which is funded by the National Aeronautics and Space Administration and operated by the California Institute of Technology, and of data, software, and web tools obtained from the High Energy Astrophysics Science Archive Research Center (HEASARC), a service of the Astrophysics Science Division at NASA/GSFC and of the Smithsonian Astrophysical Observatory's High Energy Astrophysics Division.

\appendix

\section{Model Priors} \label{apend:A}

The priors chosen for the radioactive decay model are shown in Table \ref{tab:priors_1} and those for the radioactive decay + CSM interaction model in Table \ref{tab:priors_2}. In addition, a limit on $M_{Ni}$ was given such that samples with $M_{Ni} > M_{ej}$ were rejected by the algorithm. The prior on the explosion time $t_{0}$ is based on the explosion window discussed in Section \ref{sec:obs}.

\begin{deluxetable}{cccccc}[b]
\tablecaption{Bounds for the priors used when fitting the radioactive decay model.} \label{tab:priors_1}
\tablehead{ & \colhead{$M_{ej}$} & \colhead{$M_{Ni}$} & \colhead{$v_{ej}$} &   \colhead{$\kappa_{opt}$} & \colhead{$t_{0}$}\\
& \colhead{$(M_{\odot})$} & \colhead{$(M_{\odot})$} & \colhead{$(10^3 km\,s^{-1})$} & \colhead{$(cm^2 g^{-1})$} & \colhead{($days$)}} 
\startdata
 Lower bound & $10^{-3}$ & $10^{-4}$ & 1.0 & $10^{-3}$ & -5.5\\
 Upper bound & 20.0 & 10.0 & 100.0 & 1.0 & -4.1\\
 Type & Log-uniform & Log-uniform & Uniform & Uniform & Uniform
\enddata
\end{deluxetable}

\begin{deluxetable}{cccccccccccc}[b]
\tablecaption{Bounds for the priors used when fitting the radioactive decay + CSM interaction model.} \label{tab:priors_2}
\tablehead{
 & \colhead{$M_{ej}$} & \colhead{$M_{Ni}$} & \colhead{$M_{csm}$} & \colhead{$v_{ej}$} & \colhead{$\rho_{csm, in}$} & \colhead{$r_{in}$} & \colhead{$\epsilon$} & \colhead{$x_{0}$} & \colhead{$\kappa_{opt}$} & \colhead{$t_{0}$}\\
& \colhead{$(M_{\odot})$} & \colhead{$(M_{\odot})$} & \colhead{$(M_{\odot})$} & \colhead{$(10^3 km\,s^{-1})$} & \colhead{$(10^{-12}g\,cm^{-3})$} & \colhead{$(10^{14}\,cm)$} &  &  & \colhead{$(cm^2 g^{-1})$} & \colhead{($days$)}} 
\startdata
 Lower bound & $10^{-3}$ & $10^{-4}$ & $10^{-4}$ & 1.0 & 0.01 & 0.01 & 0.1 & 0.1 & $10^{-3}$ & -5.5\\
 Upper bound & 20.0 & 10.0 & 12.0 & 20.0 & 1.0 & 20.0 & 1.0 & 1.0 & 1.0 & -4.1\\
 Type & Log-uniform & Log-uniform & Log-uniform & Uniform & Uniform & Uniform & Uniform & Uniform & Uniform & Uniform
\enddata
\end{deluxetable}

\section{Model Convergence} \label{apend:B}

For the radioactive decay + CSM interaction model fits, we use the original stopping function offered by \texttt{Dynesty}, which is robust for most applications \citep{dynesty}. The algorithm performs a ``baseline'' run, which is stopped when 99\% of the evidence has been explored, followed by a stopping function for the additional batch runs, based on whether the posterior has been estimated well enough (see \citealt{dynesty} for more details).  

\begin{figure*}
    \includegraphics[width=0.9\textwidth]{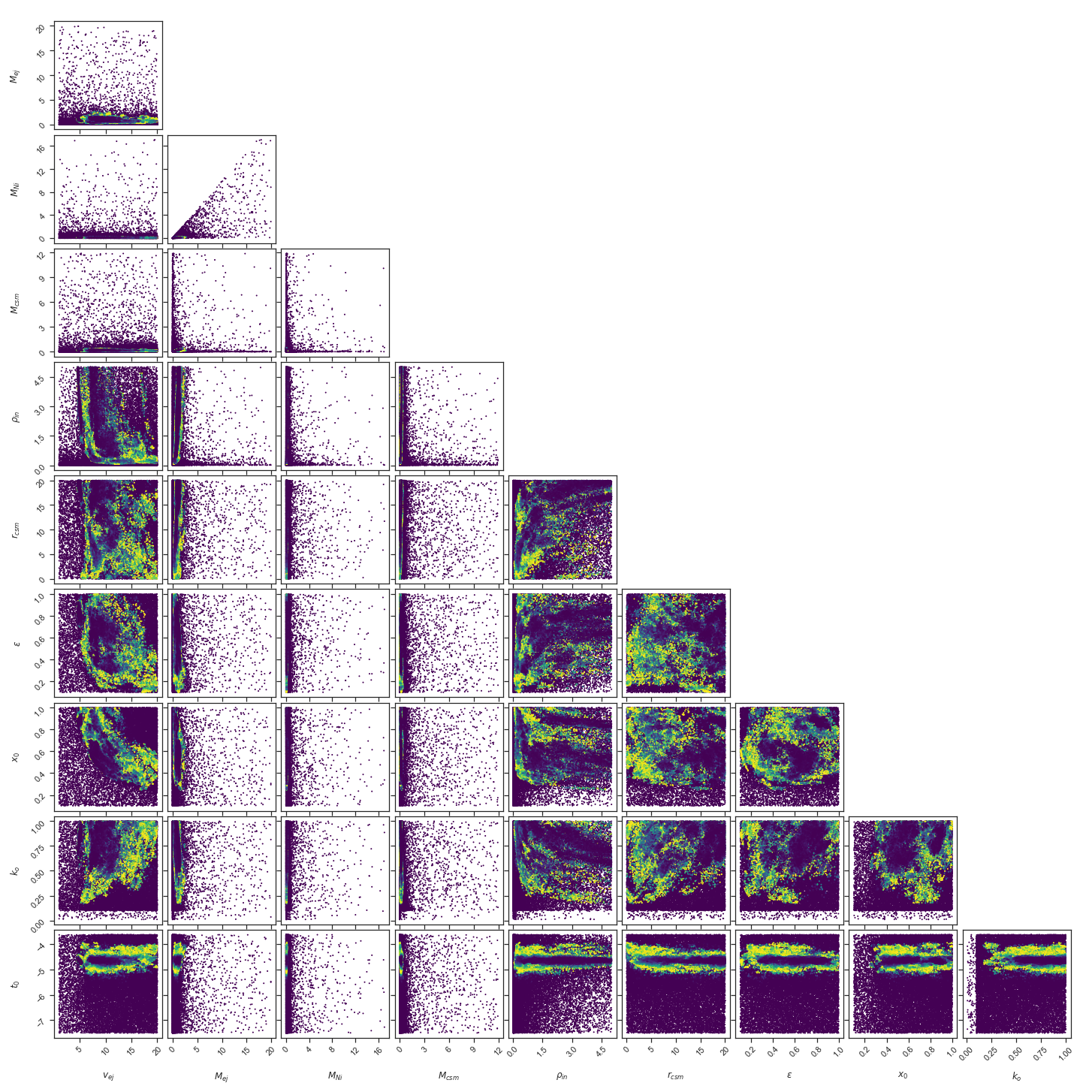}
    \caption{Corner plot of the samples constructed for the radioactive decay + CSM model with $s=0$ and $\delta=0$. Colors represent each sampling's weight in calculating the posterior. The relevant parameter space was covered well by the sampling, leading us to conclude that the fits are converged. The same is true for all other radioactive decay + CSM models fit here.}
    \label{fig:cp_points}
\end{figure*}

Taking the $s=0$, $\delta=0$ case (Fig. \ref{fig:cp_n7d0s0}) as an example, we can see two types of posterior distributions. Some (e.g. those for $M_{Ni}$ and $t_0$) are Gaussian-like, while others (e.g. for $x_0$) have broader, more complicated distributions. While this might be interpreted as not `converged' in an MCMC fit, the entire relevant phase space of the priors has been explored (Fig. \ref{fig:cp_points}). This is the case for all model variations fit here. Therefore we conclude that the fits are converged but that there exist inherently complex degeneracies between some of the parameters.

\subsection{Corner Plots and Fit Parameters} \label{apend:B1}

\begin{figure*}
    \centering
    \begin{tabular}{*{2}{@{}c}@{}}
    \includegraphics[width=0.5\textwidth]{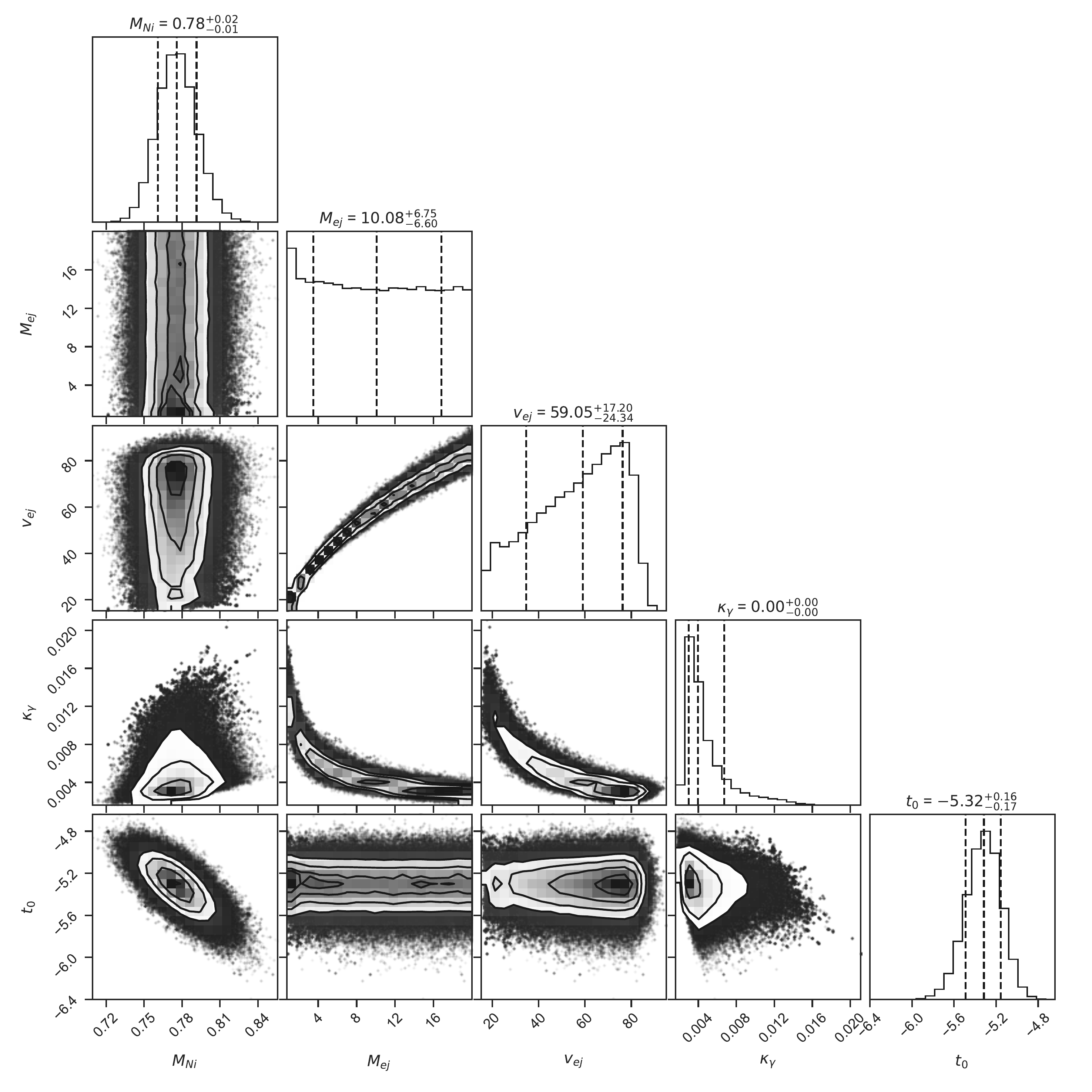}&
    \includegraphics[width=0.5\textwidth]{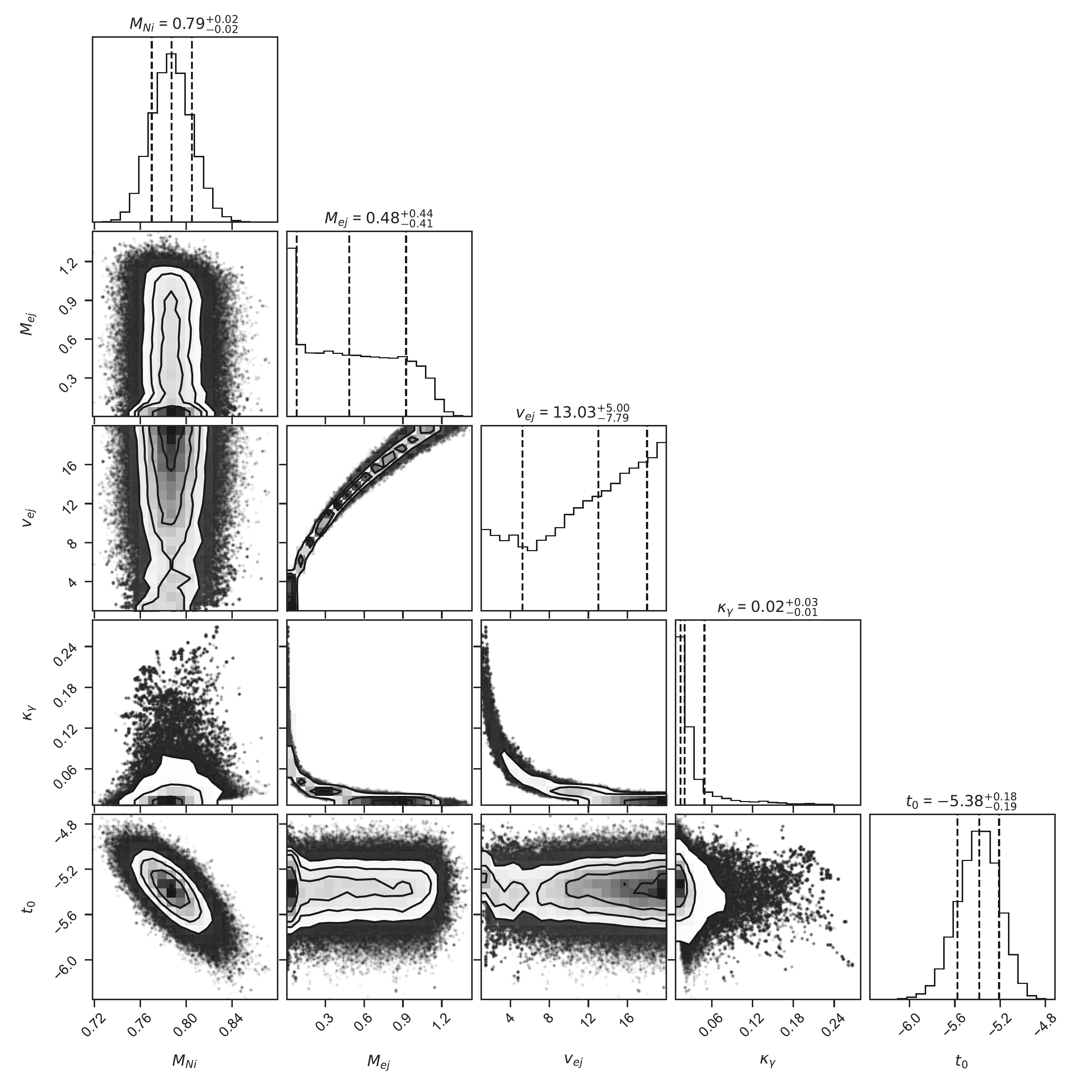}
    \end{tabular}
    \caption{Corner plots of the radioactive decay fit to the bolometric light curve of SN\,2019kbj. The units for the parameters are given in Table \ref{tab:RD_p}. Left: fit in which the $^{56}$Ni mass is constrained to be lower than the total ejecta mass and the velocity prior is extended up to $100\times10^3$\,km\,s$^{-1}$. Right: fit in which the $^{56}$Ni mass is not constrained by the ejecta mass and the velocity priors are limited to $\leq20\times10^3$\,km\,s$^{-1}$. In this case the preferred $^{56}$Ni mass is larger than the total ejecta mass, making the solution unphysical.}
    \label{fig:RD_corner}
\end{figure*}

\begin{figure*}
    \includegraphics[width=\textwidth]{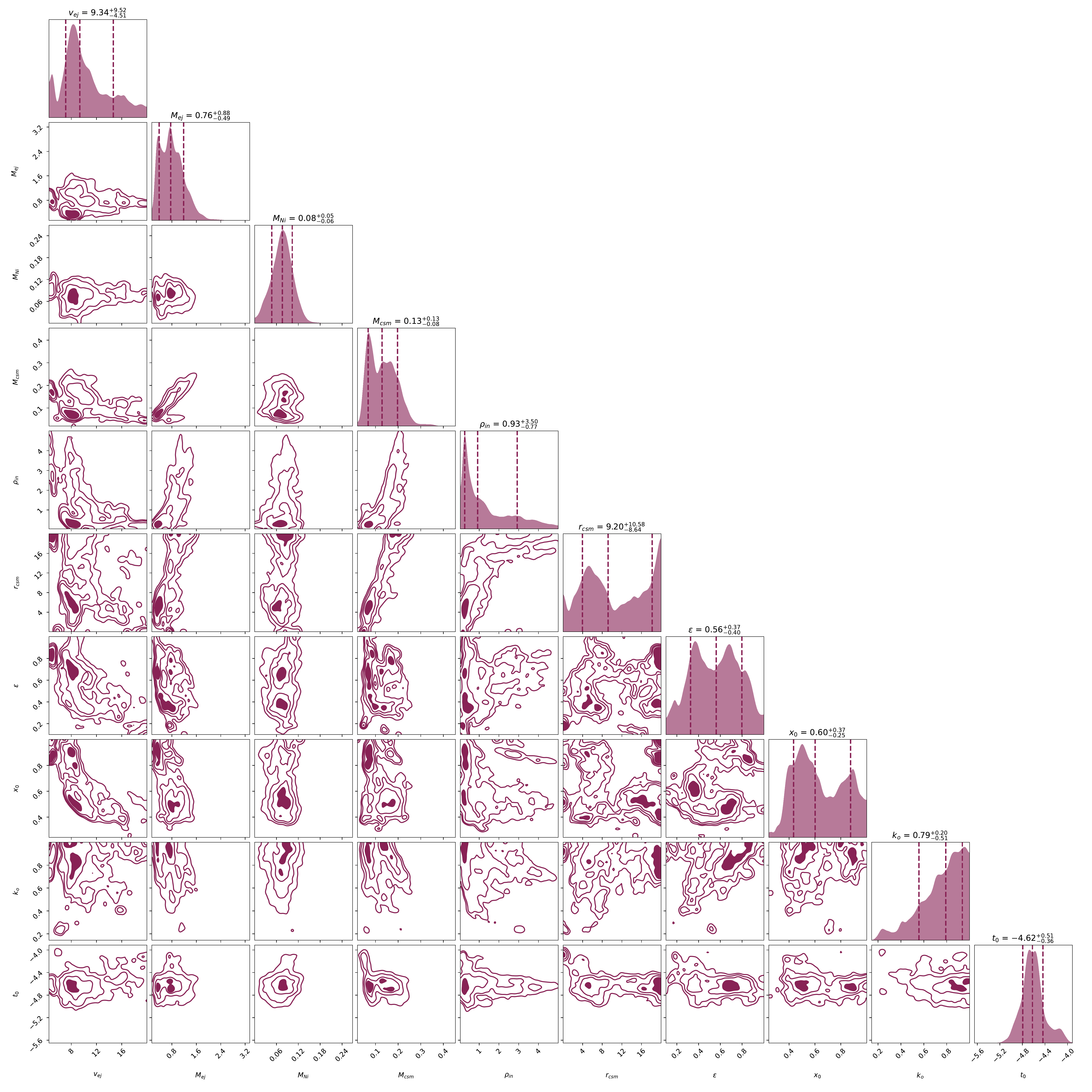}
    \caption{Corner plot for the radioactive decay + CSM interaction fit to the bolometric light curve of SN\,2019kbj, with fixed parameters $n=7, s=0, \delta=0$. The units for the parameters are given in Table \ref{tab:RDCSM_p}.} \label{fig:cp_n7d0s0}
\end{figure*}

\begin{figure*}
    \includegraphics[width=\textwidth]{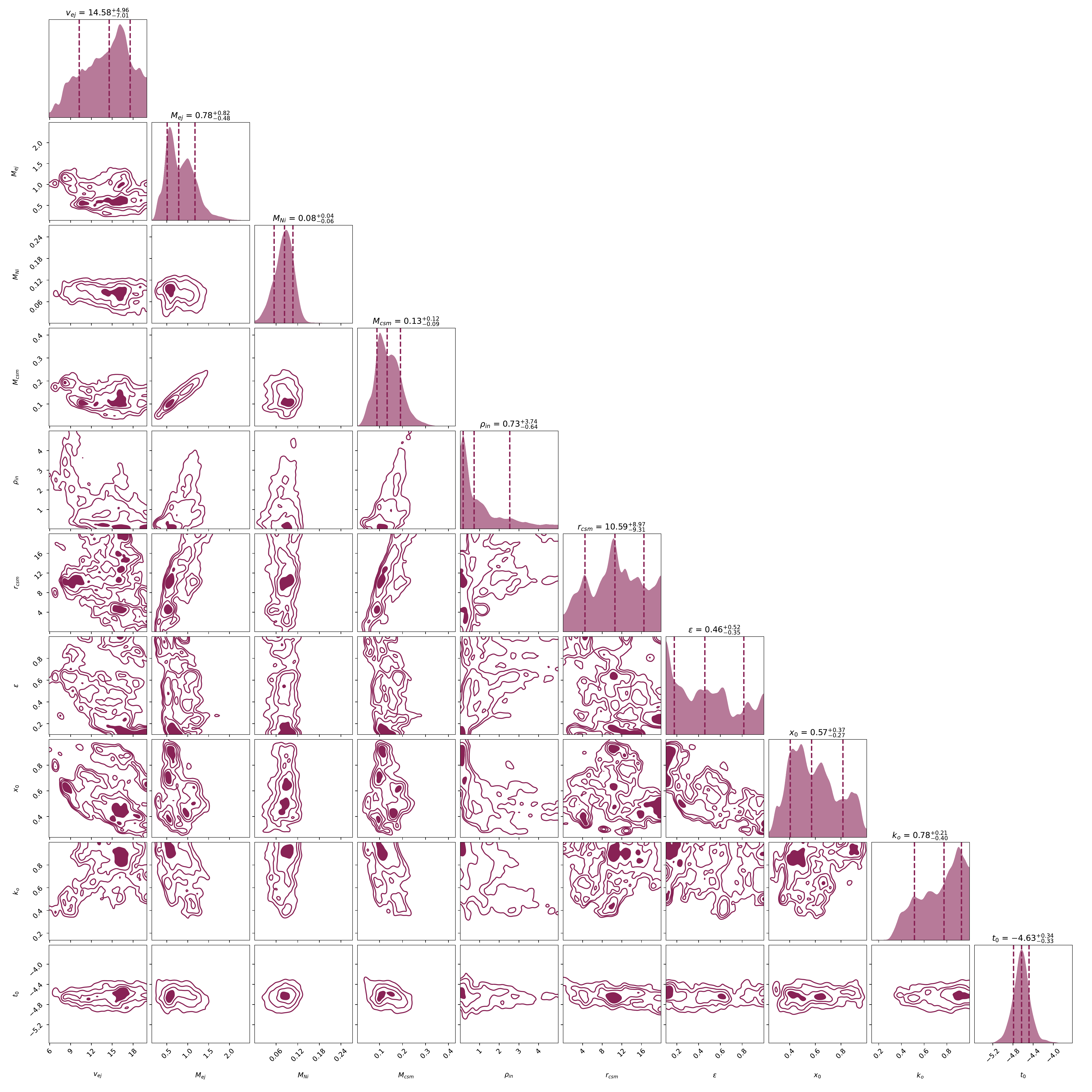}
    \caption{Corner plot for the radioactive decay + CSM interaction fit to the bolometric light curve of SN\,2019kbj, with fixed parameters $n=7, s=0, \delta=2$. The units for the parameters are given in Table \ref{tab:RDCSM_p}.} \label{fig:cp_n7d2s0}
\end{figure*}

\begin{figure*}
    \includegraphics[width=\textwidth]{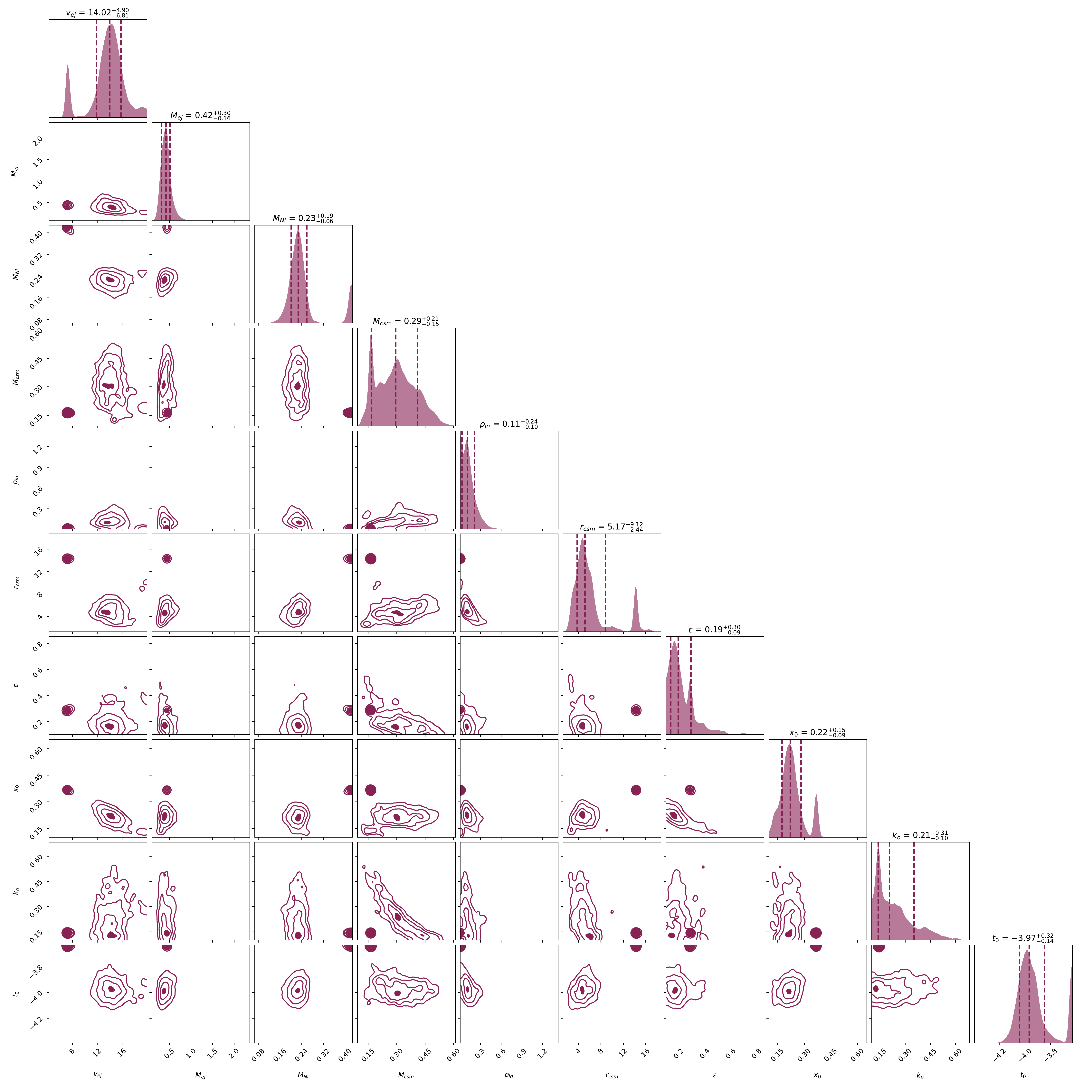}
    \caption{Corner plot for the radioactive decay + CSM interaction fit to the bolometric light curve of SN\,2019kbj, with fixed parameters $n=7, s=2, \delta=0$. The units for the parameters are given in Table \ref{tab:RDCSM_p}.} \label{fig:cp_n7d0s2}
\end{figure*}

\begin{figure*}[ht]
    \includegraphics[width=\textwidth]{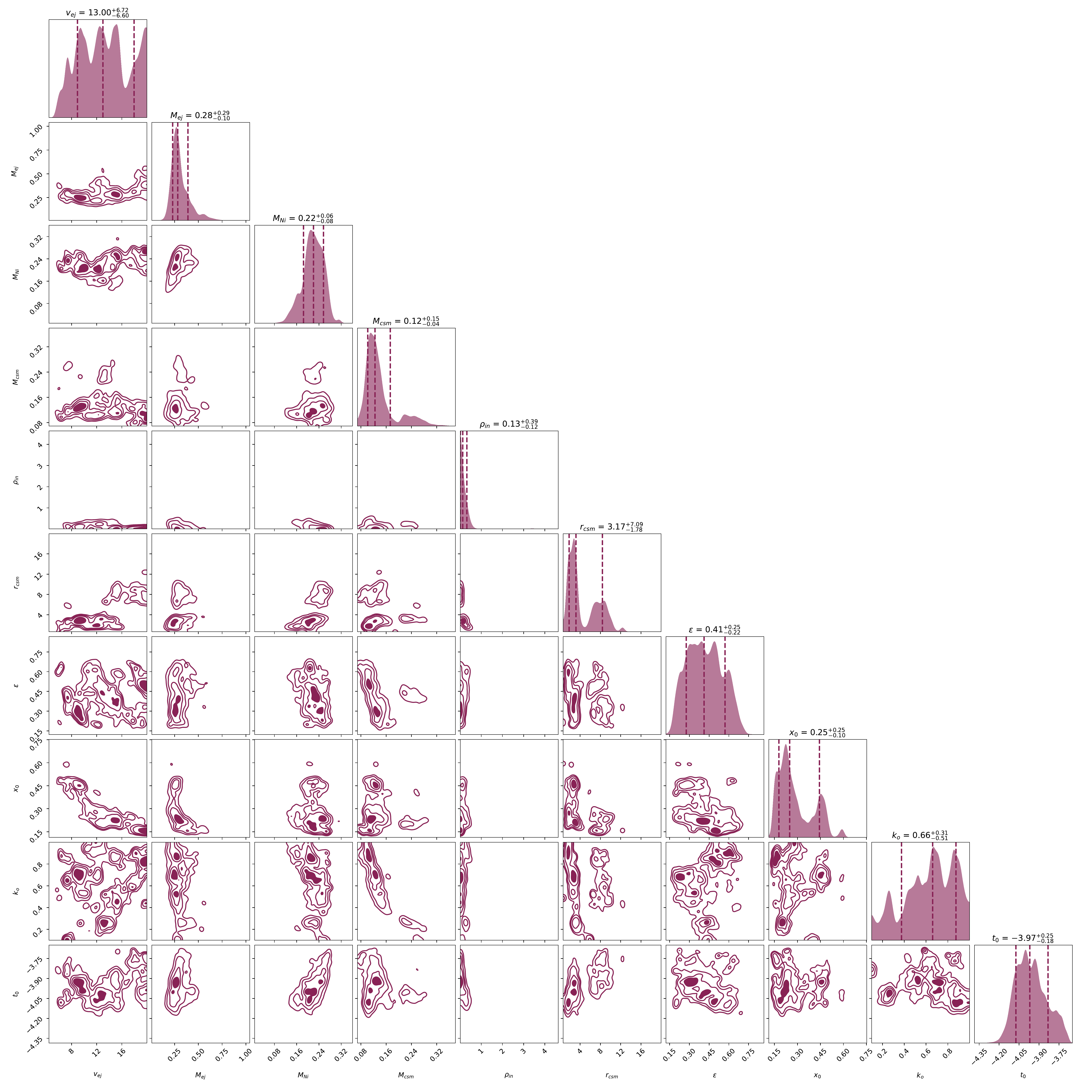}
    \caption{Corner plot for the radioactive decay + CSM interaction fit to the bolometric light curve of SN\,2019kbj, with fixed parameters $n=7, s=2, \delta=2$. The units for the parameters are given in Table \ref{tab:RDCSM_p}.} \label{fig:cp_n7d2s2}
\end{figure*}

\begin{figure*}
    \includegraphics[width=\textwidth]{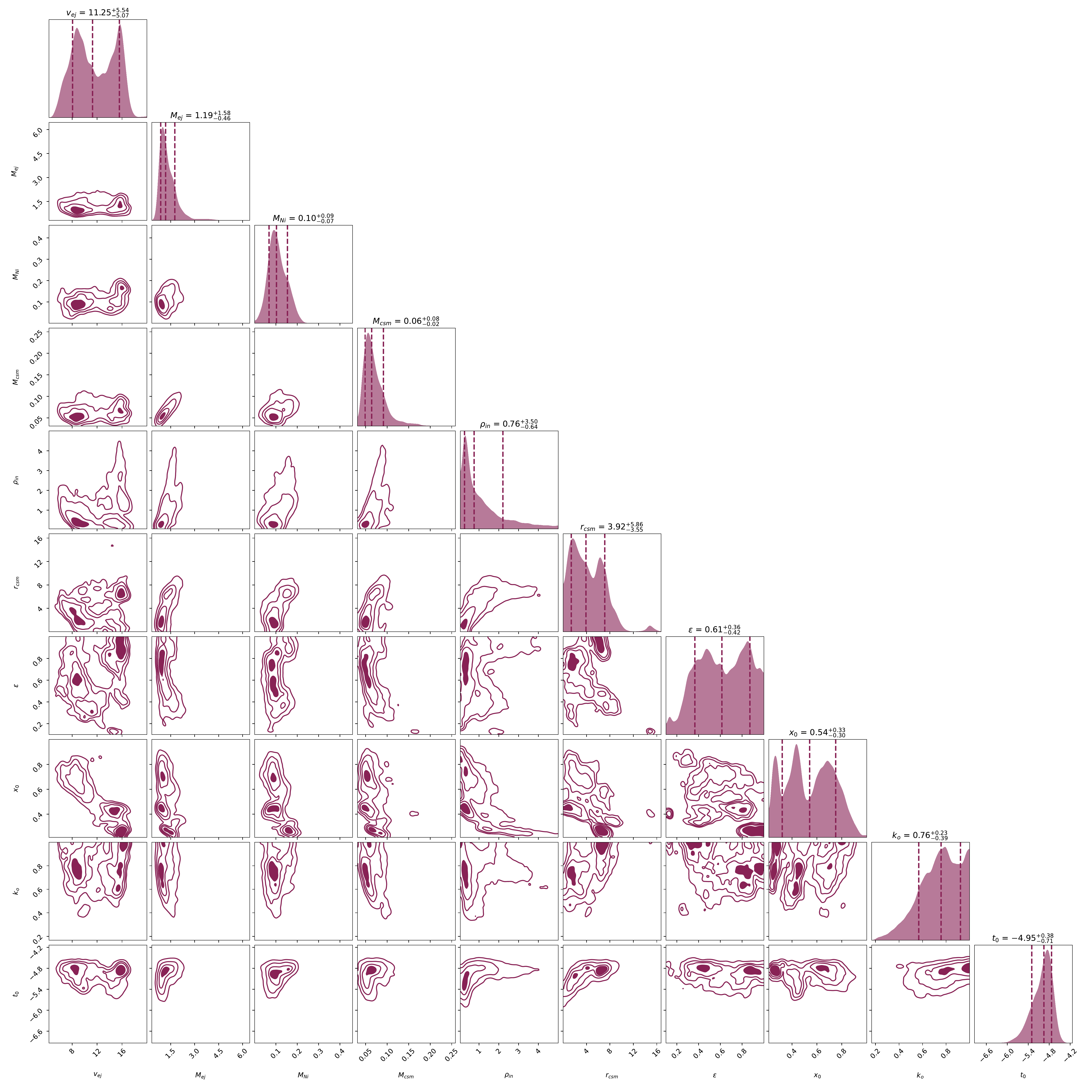}
    \caption{Corner plot for the radioactive decay + CSM interaction fit to the bolometric light curve of SN\,2019kbj, with fixed parameters $n=10, s=0, \delta=0$. The units for the parameters are given in Table \ref{tab:RDCSM_p}.} \label{fig:cp_n10d0s0}
\end{figure*}

\begin{figure*}
    \includegraphics[width=\textwidth]{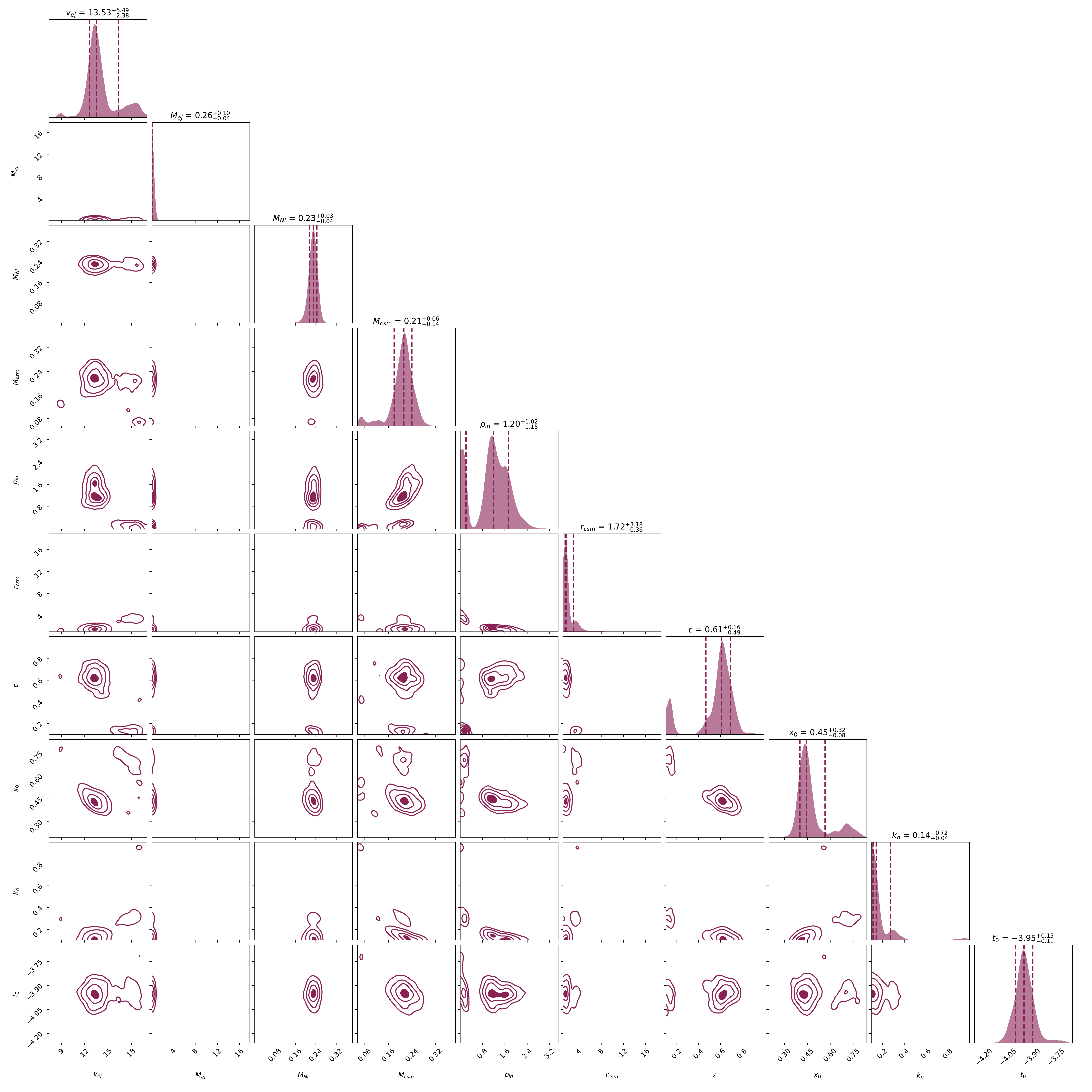}
    \caption{Corner plot for the radioactive decay + CSM interaction fit to the bolometric light curve of SN\,2019kbj, with fixed parameters $n=10, s=0, \delta=2$. The units for the parameters are given in Table \ref{tab:RDCSM_p}.} \label{fig:cp_n10d2s0}
\end{figure*}

\begin{figure*}
    \includegraphics[width=\textwidth]{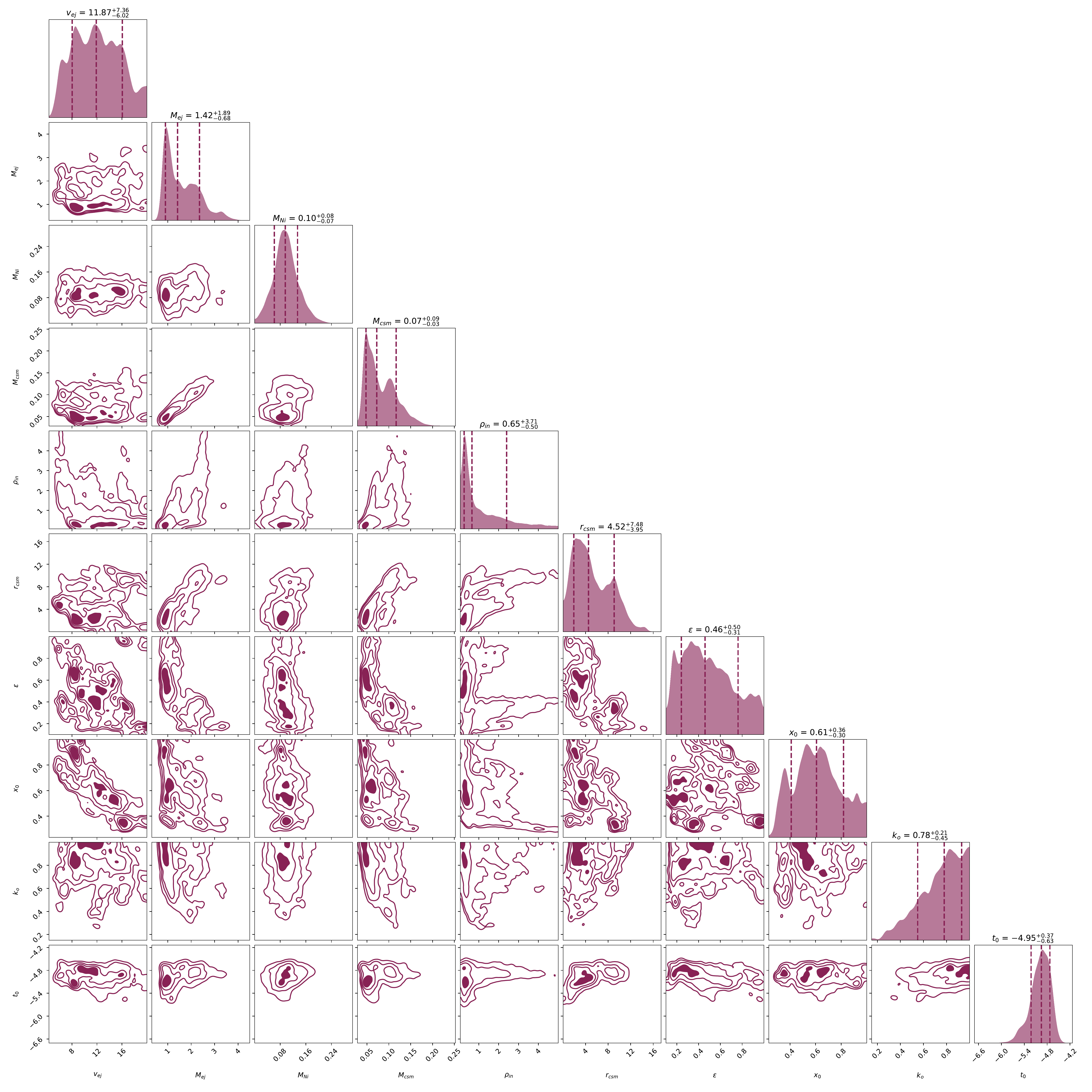}
    \caption{Corner plot for the radioactive decay + CSM interaction fit to the bolometric light curve of SN\,2019kbj, with fixed parameters $n=10, s=2, \delta=0$. The units for the parameters are given in Table \ref{tab:RDCSM_p}.} \label{fig:cp_n10d0s2}
\end{figure*}

\begin{figure*}[ht]
    \includegraphics[width=\textwidth]{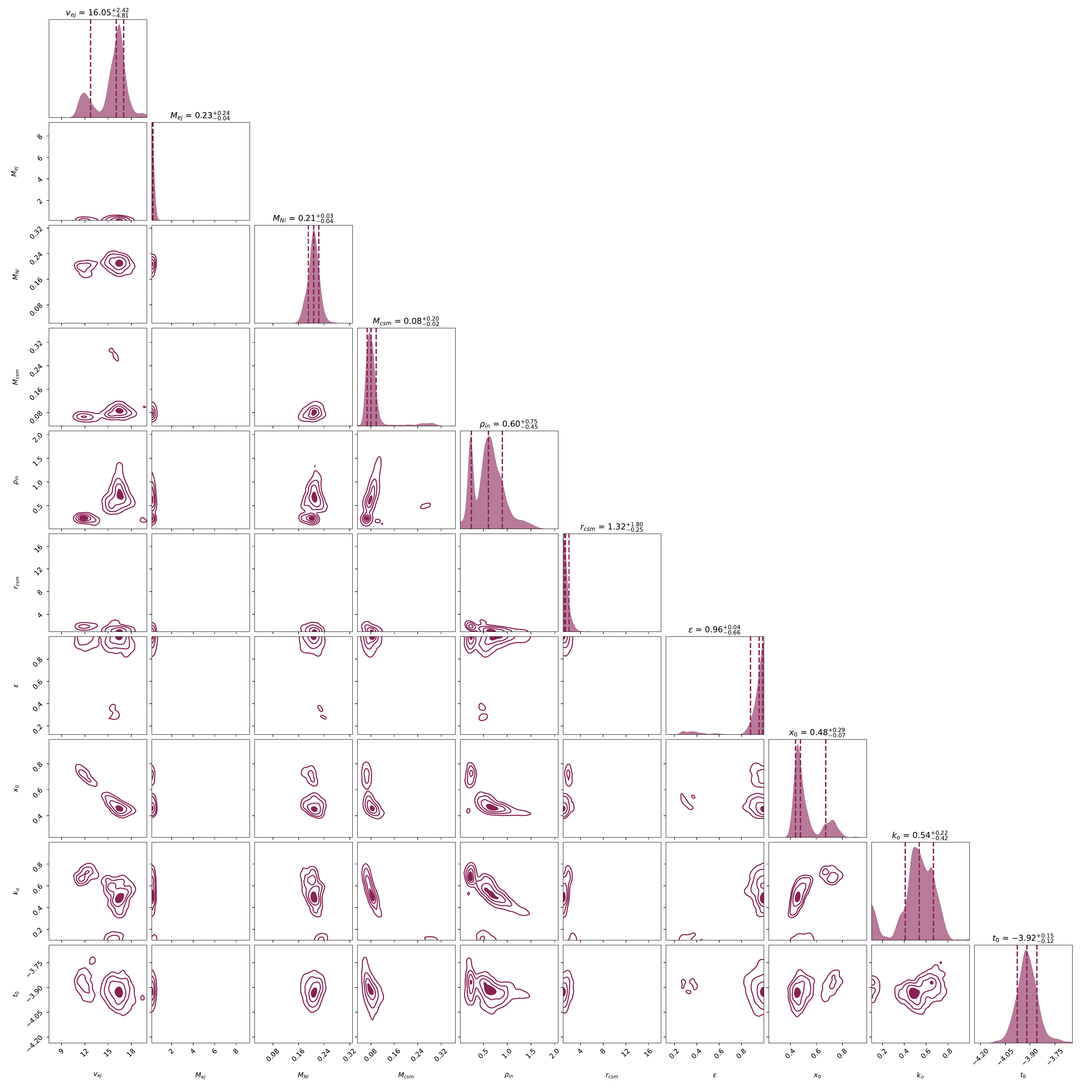}
    \caption{Corner plot for the radioactive decay + CSM interaction fit to the bolometric light curve of SN\,2019kbj, with fixed parameters $n=10, s=2, \delta=2$. The units for the parameters are given in Table \ref{tab:RDCSM_p}.} \label{fig:cp_n10d2s2}
\end{figure*}

Figure \ref{fig:RD_corner} shows the corner plot for the radioactive decay only fit, and Table \ref{tab:uc_RD_p} lists the best-fit parameters from the unconstrained model. Corner plots for the radioactive decay + CSM fits are shown in Figures \ref{fig:cp_n7d0s0}--\ref{fig:cp_n10d2s2}.

\begin{deluxetable}{ccccc}[hb]
\tabletypesize{\footnotesize}
\tablecaption{Best-fit parameters for the unconstrained radioactive decay model. The larger Ni than total mass makes this model unphysical.} \label{tab:uc_RD_p}
\tablehead{
\colhead{$M_{Ni}$} & \colhead{$M_{ej}$} & \colhead{$v_{ej}$} & \colhead{$\kappa_{opt}$} & \colhead{$t_{0}$}\\
\colhead{($M_{\odot}$)} & \colhead{($M_{\odot}$)} & \colhead{($10^3 km\,s^{-1}$)} & \colhead{($cm^2 g^{-1}$)} & \colhead{($days$)}}
\startdata
$0.787_{-0.017}^{+0.018}$ & $0.48_{-0.41}^{+0.44}$ & $13.02_{-7.79}^{+5.00}$ & $0.020_{-0.006}^{+0.029}$ & $-5.38_{-0.19}^{+0.18}$ \\
\enddata
\end{deluxetable}

\bibliography{References}{}

\begin{thebibliography}{}
\expandafter\ifx\csname natexlab\endcsname\relax\def\natexlab#1{#1}\fi
\providecommand{\url}[1]{\href{#1}{#1}}
\providecommand{\dodoi}[1]{doi:~\href{http://doi.org/#1}{\nolinkurl{#1}}}
\providecommand{\doeprint}[1]{\href{http://ascl.net/#1}{\nolinkurl{http://ascl.net/#1}}}
\providecommand{\doarXiv}[1]{\href{https://arxiv.org/abs/#1}{\nolinkurl{https://arxiv.org/abs/#1}}}

\bibitem[{{Alard} \& {Lupton}(1998)}]{Alard1998}
{Alard}, C., \& {Lupton}, R.~H. 1998, \apj, 503, 325, \dodoi{10.1086/305984}

\bibitem[{{Arcavi}(2022)}]{Arcavi2022blackbodies}
{Arcavi}, I. 2022, \apj, 937, 75, \dodoi{10.3847/1538-4357/ac90c0}

\bibitem[{{Arcavi} {et~al.}(2022){Arcavi}, {Ben-Ami}, {Hiramatsu}, {Howell},
  {Burke}, {Pellegrino}, \& {McCully}}]{Arcavi2022}
{Arcavi}, I., {Ben-Ami}, T., {Hiramatsu}, D., {et~al.} 2022, Transient Name
  Server Classification Report, 2022-33, 1

\bibitem[{Arnett(1982)}]{arnett82}
Arnett, W.~D. 1982, The Astrophysical Journal, 253, 785

\bibitem[{{Becker}(2015)}]{Becker2015}
{Becker}, A. 2015, {HOTPANTS: High Order Transform of PSF ANd Template
  Subtraction}, Astrophysics Source Code Library, record ascl:1504.004.
\newblock \doeprint{1504.004}

\bibitem[{Ben-Ami(2022)}]{bolmod}
Ben-Ami, T. 2022, \dodoi{10.5281/zenodo.5834661}

\bibitem[{{Brown} {et~al.}(2014){Brown}, {Breeveld}, {Holland}, {Kuin}, \&
  {Pritchard}}]{Brown14}
{Brown}, P.~J., {Breeveld}, A.~A., {Holland}, S., {Kuin}, P., \& {Pritchard},
  T. 2014, \apss, 354, 89, \dodoi{10.1007/s10509-014-2059-8}

\bibitem[{{Brown} {et~al.}(2009){Brown}, {Holland}, {Immler}, {Milne},
  {Roming}, {Gehrels}, {Nousek}, {Panagia}, {Still}, \& {Vanden
  Berk}}]{Brown09}
{Brown}, P.~J., {Holland}, S.~T., {Immler}, S., {et~al.} 2009, \aj, 137, 4517,
  \dodoi{10.1088/0004-6256/137/5/4517}

\bibitem[{{Brown} {et~al.}(2013){Brown}, {Baliber}, {Bianco}, {Bowman},
  {Burleson}, {Conway}, {Crellin}, {Depagne}, {De Vera}, {Dilday}, {Dragomir},
  {Dubberley}, {Eastman}, {Elphick}, {Falarski}, {Foale}, {Ford}, {Fulton},
  {Garza}, {Gomez}, {Graham}, {Greene}, {Haldeman}, {Hawkins}, {Haworth},
  {Haynes}, {Hidas}, {Hjelstrom}, {Howell}, {Hygelund}, {Lister}, {Lobdill},
  {Martinez}, {Mullins}, {Norbury}, {Parrent}, {Paulson}, {Petry}, {Pickles},
  {Posner}, {Rosing}, {Ross}, {Sand}, {Saunders}, {Shobbrook}, {Shporer},
  {Street}, {Thomas}, {Tsapras}, {Tufts}, {Valenti}, {Vander Horst}, {Walker},
  {White}, \& {Willis}}]{Brown2013}
{Brown}, T.~M., {Baliber}, N., {Bianco}, F.~B., {et~al.} 2013, \pasp, 125,
  1031, \dodoi{10.1086/673168}

\bibitem[{Cappellaro {et~al.}(1997)Cappellaro, Mazzali, Benetti, Danziger,
  Turatto, Della~Valle, \& Patat}]{cappellaro97}
Cappellaro, E., Mazzali, P., Benetti, S., {et~al.} 1997, arXiv preprint
  astro-ph/9707016

\bibitem[{{Chambers} {et~al.}(2016){Chambers}, {Magnier}, {Metcalfe},
  {Flewelling}, {Huber}, {Waters}, {Denneau}, {Draper}, {Farrow}, {Finkbeiner},
  {Holmberg}, {Koppenhoefer}, {Price}, {Rest}, {Saglia}, {Schlafly}, {Smartt},
  {Sweeney}, {Wainscoat}, {Burgett}, {Chastel}, {Grav}, {Heasley}, {Hodapp},
  {Jedicke}, {Kaiser}, {Kudritzki}, {Luppino}, {Lupton}, {Monet}, {Morgan},
  {Onaka}, {Shiao}, {Stubbs}, {Tonry}, {White}, {Ba{\~n}ados}, {Bell},
  {Bender}, {Bernard}, {Boegner}, {Boffi}, {Botticella}, {Calamida},
  {Casertano}, {Chen}, {Chen}, {Cole}, {Deacon}, {Frenk}, {Fitzsimmons},
  {Gezari}, {Gibbs}, {Goessl}, {Goggia}, {Gourgue}, {Goldman}, {Grant},
  {Grebel}, {Hambly}, {Hasinger}, {Heavens}, {Heckman}, {Henderson}, {Henning},
  {Holman}, {Hopp}, {Ip}, {Isani}, {Jackson}, {Keyes}, {Koekemoer}, {Kotak},
  {Le}, {Liska}, {Long}, {Lucey}, {Liu}, {Martin}, {Masci}, {McLean}, {Mindel},
  {Misra}, {Morganson}, {Murphy}, {Obaika}, {Narayan}, {Nieto-Santisteban},
  {Norberg}, {Peacock}, {Pier}, {Postman}, {Primak}, {Rae}, {Rai}, {Riess},
  {Riffeser}, {Rix}, {R{\"o}ser}, {Russel}, {Rutz}, {Schilbach}, {Schultz},
  {Scolnic}, {Strolger}, {Szalay}, {Seitz}, {Small}, {Smith}, {Soderblom},
  {Taylor}, {Thomson}, {Taylor}, {Thakar}, {Thiel}, {Thilker}, {Unger},
  {Urata}, {Valenti}, {Wagner}, {Walder}, {Walter}, {Watters}, {Werner},
  {Wood-Vasey}, \& {Wyse}}]{Chambers2016}
{Chambers}, K.~C., {Magnier}, E.~A., {Metcalfe}, N., {et~al.} 2016, arXiv
  e-prints, arXiv:1612.05560.
\newblock \doarXiv{1612.05560}

\bibitem[{Chatzopoulos {et~al.}(2012)Chatzopoulos, Wheeler, \&
  Vinko}]{chatzopoulos12}
Chatzopoulos, E., Wheeler, J.~C., \& Vinko, J. 2012, The Astrophysical Journal,
  746, 121

\bibitem[{{Chevalier}(1981)}]{Chevalier1981}
{Chevalier}, R.~A. 1981, \apj, 246, 267, \dodoi{10.1086/158920}

\bibitem[{Chevalier(1982)}]{chevalier82}
Chevalier, R.~A. 1982, The Astrophysical Journal, 258, 790

\bibitem[{Crowther(2007)}]{crowther07}
Crowther, P.~A. 2007, Annu. Rev. Astron. Astrophys., 45, 177

\bibitem[{Filippenko(1982)}]{Filippenko_1982}
Filippenko, A.~V. 1982, Publications of the Astronomical Society of the
  Pacific, 94, 715, \dodoi{10.1086/131052}

\bibitem[{{Flewelling} {et~al.}(2020){Flewelling}, {Magnier}, {Chambers},
  {Heasley}, {Holmberg}, {Huber}, {Sweeney}, {Waters}, {Calamida}, {Casertano},
  {Chen}, {Farrow}, {Hasinger}, {Henderson}, {Long}, {Metcalfe}, {Narayan},
  {Nieto-Santisteban}, {Norberg}, {Rest}, {Saglia}, {Szalay}, {Thakar},
  {Tonry}, {Valenti}, {Werner}, {White}, {Denneau}, {Draper}, {Hodapp},
  {Jedicke}, {Kaiser}, {Kudritzki}, {Price}, {Wainscoat}, {Chastel}, {McLean},
  {Postman}, \& {Shiao}}]{Flewelling2020}
{Flewelling}, H.~A., {Magnier}, E.~A., {Chambers}, K.~C., {et~al.} 2020, \apjs,
  251, 7, \dodoi{10.3847/1538-4365/abb82d}

\bibitem[{Foley {et~al.}(2015)Foley, Zheng, Filippenko, \& Dyk}]{foley15}
Foley, R.~J., Zheng, W., Filippenko, A.~V., \& Dyk, S. D.~V. 2015, ATel, 7298,
  1

\bibitem[{Foreman-Mackey {et~al.}(2013)Foreman-Mackey, Hogg, Lang, \&
  Goodman}]{emcee}
Foreman-Mackey, D., Hogg, D.~W., Lang, D., \& Goodman, J. 2013, Publications of
  the Astronomical Society of the Pacific, 125, 306

\bibitem[{Gangopadhyay {et~al.}(2020)Gangopadhyay, Misra, Hiramatsu, Wang, \&
  Hosseinzadeh}]{Gangopadhyay_2020}
Gangopadhyay, A., Misra, K., Hiramatsu, D., Wang, S.-Q., \& Hosseinzadeh, G.
  2020, The Astrophysical Journal, 889, 170, \dodoi{10.3847/1538-4357/ab6328}

\bibitem[{{Gehrels} {et~al.}(2004){Gehrels}, {Chincarini}, {Giommi}, {Mason},
  {Nousek}, {Wells}, {White}, {Barthelmy}, {Burrows}, {Cominsky}, {Hurley},
  {Marshall}, {M{\'e}sz{\'a}ros}, {Roming}, {Angelini}, {Barbier}, {Belloni},
  {Campana}, {Caraveo}, {Chester}, {Citterio}, {Cline}, {Cropper}, {Cummings},
  {Dean}, {Feigelson}, {Fenimore}, {Frail}, {Fruchter}, {Garmire}, {Gendreau},
  {Ghisellini}, {Greiner}, {Hill}, {Hunsberger}, {Krimm}, {Kulkarni}, {Kumar},
  {Lebrun}, {Lloyd-Ronning}, {Markwardt}, {Mattson}, {Mushotzky}, {Norris},
  {Osborne}, {Paczynski}, {Palmer}, {Park}, {Parsons}, {Paul}, {Rees},
  {Reynolds}, {Rhoads}, {Sasseen}, {Schaefer}, {Short}, {Smale}, {Smith},
  {Stella}, {Tagliaferri}, {Takahashi}, {Tashiro}, {Townsley}, {Tueller},
  {Turner}, {Vietri}, {Voges}, {Ward}, {Willingale}, {Zerbi}, \&
  {Zhang}}]{Gehrels04}
{Gehrels}, N., {Chincarini}, G., {Giommi}, P., {et~al.} 2004, \apj, 611, 1005,
  \dodoi{10.1086/422091}

\bibitem[{{Hiramatsu} {et~al.}(2019){Hiramatsu}, {Arcavi}, {Burke},
  {Hosseinzadeh}, {Howell}, {McCully}, {Pellegrino}, \&
  {Valenti}}]{classification}
{Hiramatsu}, D., {Arcavi}, I., {Burke}, J., {et~al.} 2019, Transient Name
  Server Classification Report, 2019-1137, 1

\bibitem[{Hosseinzadeh {et~al.}(2017)Hosseinzadeh, Arcavi, Valenti, McCully,
  Howell, \& Johansson}]{Hosseinzadeh_2017}
Hosseinzadeh, G., Arcavi, I., Valenti, S., {et~al.} 2017, The Astrophysical
  Journal, 836, 158, \dodoi{10.3847/1538-4357/836/2/158}

\bibitem[{{Hosseinzadeh} \& {Gomez}(2020)}]{lc_fit}
{Hosseinzadeh}, G., \& {Gomez}, S. 2020, Light Curve Fitting, v0.2.0,  Zenodo,
  \dodoi{10.5281/zenodo.4312178}

\bibitem[{{Hosseinzadeh} {et~al.}(2019){Hosseinzadeh}, {McCully}, {Zabludoff},
  {Arcavi}, {French}, {Howell}, {Berger}, \& {Hiramatsu}}]{Hosseinzadeh2019}
{Hosseinzadeh}, G., {McCully}, C., {Zabludoff}, A.~I., {et~al.} 2019, \apjl,
  871, L9, \dodoi{10.3847/2041-8213/aafc61}

\bibitem[{{Karamehmetoglu} {et~al.}(2017){Karamehmetoglu}, {Taddia},
  {Sollerman}, {Wyrzykowski}, {Schmidl}, {Fraser}, {Fremling}, {Greiner},
  {Inserra}, {Kostrzewa-Rutkowska}, {Maguire}, {Smartt}, {Sullivan}, \&
  {Young}}]{Karamehmetoglu2017}
{Karamehmetoglu}, E., {Taddia}, F., {Sollerman}, J., {et~al.} 2017, \aap, 602,
  A93, \dodoi{10.1051/0004-6361/201629619}

\bibitem[{{Khazov} {et~al.}(2016){Khazov}, {Yaron}, {Gal-Yam}, {Manulis},
  {Rubin}, {Kulkarni}, {Arcavi}, {Kasliwal}, {Ofek}, {Cao}, {Perley},
  {Sollerman}, {Horesh}, {Sullivan}, {Filippenko}, {Nugent}, {Howell}, {Cenko},
  {Silverman}, {Ebeling}, {Taddia}, {Johansson}, {Laher}, {Surace},
  {Rebbapragada}, {Wozniak}, \& {Matheson}}]{Khazov2016}
{Khazov}, D., {Yaron}, O., {Gal-Yam}, A., {et~al.} 2016, \apj, 818, 3,
  \dodoi{10.3847/0004-637X/818/1/3}

\bibitem[{{Kiewe} {et~al.}(2012){Kiewe}, {Gal-Yam}, {Arcavi}, {Leonard},
  {Emilio Enriquez}, {Cenko}, {Fox}, {Moon}, {Sand}, {Soderberg}, \&
  {CCCP}}]{Kiewe2012}
{Kiewe}, M., {Gal-Yam}, A., {Arcavi}, I., {et~al.} 2012, \apj, 744, 10,
  \dodoi{10.1088/0004-637X/744/1/10}

\bibitem[{{Matzner} \& {McKee}(1999)}]{Matzner1999}
{Matzner}, C.~D., \& {McKee}, C.~F. 1999, \apj, 510, 379,
  \dodoi{10.1086/306571}

\bibitem[{{McCully} {et~al.}(2018){McCully}, {Volgenau}, {Harbeck}, {Lister},
  {Saunders}, {Turner}, {Siiverd}, \& {Bowman}}]{McCully2018}
{McCully}, C., {Volgenau}, N.~H., {Harbeck}, D.-R., {et~al.} 2018, in Society
  of Photo-Optical Instrumentation Engineers (SPIE) Conference Series, Vol.
  10707, Software and Cyberinfrastructure for Astronomy V, ed. J.~C. {Guzman}
  \& J.~{Ibsen}, 107070K, \dodoi{10.1117/12.2314340}

\bibitem[{Nadyozhin(1994)}]{decayprop}
Nadyozhin, D. 1994, The Astrophysical Journal Supplement Series, 92, 527

\bibitem[{{Nyholm} {et~al.}(2017){Nyholm}, {Sollerman}, {Taddia}, {Fremling},
  {Moriya}, {Ofek}, {Gal-Yam}, {De Cia}, {Roy}, {Kasliwal}, {Cao}, {Nugent}, \&
  {Masci}}]{Nyholm2017}
{Nyholm}, A., {Sollerman}, J., {Taddia}, F., {et~al.} 2017, \aap, 605, A6,
  \dodoi{10.1051/0004-6361/201629906}

\bibitem[{Pastorello {et~al.}(2015{\natexlab{a}})Pastorello, Benetti, Brown,
  Tsvetkov, Inserra, \& Taubenberger}]{pastorello15a}
Pastorello, A., Benetti, S., Brown, P.~J., {et~al.} 2015{\natexlab{a}}, Monthly
  Notices of the Royal Astronomical Society, 449, 1921,
  \dodoi{10.1093/mnras/stu2745}

\bibitem[{Pastorello {et~al.}(2007)Pastorello, Smartt, Mattila, Eldridge,
  Young, Itagaki, Yamaoka, Navasardyan, Valenti, Patat,
  {et~al.}}]{pastorello2007}
Pastorello, A., Smartt, S., Mattila, S., {et~al.} 2007, Nature, 447, 829

\bibitem[{Pastorello {et~al.}(2008{\natexlab{a}})Pastorello, Mattila, Zampieri,
  Della~Valle, Smartt, Valenti, Agnoletto, Benetti, Benn, Branch,
  {et~al.}}]{pastorello08}
Pastorello, A., Mattila, S., Zampieri, L., {et~al.} 2008{\natexlab{a}}, Monthly
  Notices of the Royal Astronomical Society, 389, 113

\bibitem[{Pastorello {et~al.}(2008{\natexlab{b}})Pastorello, Quimby, Smartt,
  Mattila, Navasardyan, Crockett, Elias-Rosa, Mondol, Wheeler, \&
  Young}]{pastorello08b}
Pastorello, A., Quimby, R., Smartt, S., {et~al.} 2008{\natexlab{b}}, Monthly
  Notices of the Royal Astronomical Society, 389, 131

\bibitem[{Pastorello {et~al.}(2015{\natexlab{b}})Pastorello, Wyrzykowski,
  Valenti, Prieto, Koz{\l}owski, Udalski, Elias-Rosa, Morales-Garoffolo,
  Anderson, Benetti, {et~al.}}]{pastorello15e}
Pastorello, A., Wyrzykowski, {\L}., Valenti, S., {et~al.} 2015{\natexlab{b}},
  Monthly Notices of the Royal Astronomical Society, 449, 1941

\bibitem[{Pastorello {et~al.}(2016)Pastorello, Wang, Ciabattari, Bersier,
  Mazzali, Gao, Xu, Zhang, Tokuoka, Benetti, {et~al.}}]{pastorello16}
Pastorello, A., Wang, X.-F., Ciabattari, F., {et~al.} 2016, Monthly Notices of
  the Royal Astronomical Society, 456, 853

\bibitem[{{Pellegrino} {et~al.}(2022){Pellegrino}, {Howell}, {Vink{\'o}},
  {Gangopadhyay}, {Xiang}, {Arcavi}, {Brown}, {Burke}, {Hiramatsu},
  {Hosseinzadeh}, {Li}, {McCully}, {Misra}, {Newsome}, {Gonzalez}, {Pritchard},
  {Valenti}, {Wang}, \& {Zhang}}]{Pellegrino2022}
{Pellegrino}, C., {Howell}, D.~A., {Vink{\'o}}, J., {et~al.} 2022, \apj, 926,
  125, \dodoi{10.3847/1538-4357/ac3e63}

\bibitem[{{Planck Collaboration} {et~al.}(2020){Planck Collaboration},
  {Aghanim}, {Akrami}, {Ashdown}, {Aumont}, {Baccigalupi}, {Ballardini},
  {Banday}, {Barreiro}, {Bartolo}, {Basak}, {Battye}, {Benabed}, {Bernard},
  {Bersanelli}, {Bielewicz}, {Bock}, {Bond}, {Borrill}, {Bouchet}, {Boulanger},
  {Bucher}, {Burigana}, {Butler}, {Calabrese}, {Cardoso}, {Carron},
  {Challinor}, {Chiang}, {Chluba}, {Colombo}, {Combet}, {Contreras}, {Crill},
  {Cuttaia}, {de Bernardis}, {de Zotti}, {Delabrouille}, {Delouis}, {Di
  Valentino}, {Diego}, {Dor{\'e}}, {Douspis}, {Ducout}, {Dupac}, {Dusini},
  {Efstathiou}, {Elsner}, {En{\ss}lin}, {Eriksen}, {Fantaye}, {Farhang},
  {Fergusson}, {Fernandez-Cobos}, {Finelli}, {Forastieri}, {Frailis},
  {Fraisse}, {Franceschi}, {Frolov}, {Galeotta}, {Galli}, {Ganga},
  {G{\'e}nova-Santos}, {Gerbino}, {Ghosh}, {Gonz{\'a}lez-Nuevo}, {G{\'o}rski},
  {Gratton}, {Gruppuso}, {Gudmundsson}, {Hamann}, {Handley}, {Hansen},
  {Herranz}, {Hildebrandt}, {Hivon}, {Huang}, {Jaffe}, {Jones}, {Karakci},
  {Keih{\"a}nen}, {Keskitalo}, {Kiiveri}, {Kim}, {Kisner}, {Knox},
  {Krachmalnicoff}, {Kunz}, {Kurki-Suonio}, {Lagache}, {Lamarre}, {Lasenby},
  {Lattanzi}, {Lawrence}, {Le Jeune}, {Lemos}, {Lesgourgues}, {Levrier},
  {Lewis}, {Liguori}, {Lilje}, {Lilley}, {Lindholm}, {L{\'o}pez-Caniego},
  {Lubin}, {Ma}, {Mac{\'\i}as-P{\'e}rez}, {Maggio}, {Maino}, {Mandolesi},
  {Mangilli}, {Marcos-Caballero}, {Maris}, {Martin}, {Martinelli},
  {Mart{\'\i}nez-Gonz{\'a}lez}, {Matarrese}, {Mauri}, {McEwen}, {Meinhold},
  {Melchiorri}, {Mennella}, {Migliaccio}, {Millea}, {Mitra},
  {Miville-Desch{\^e}nes}, {Molinari}, {Montier}, {Morgante}, {Moss}, {Natoli},
  {N{\o}rgaard-Nielsen}, {Pagano}, {Paoletti}, {Partridge}, {Patanchon},
  {Peiris}, {Perrotta}, {Pettorino}, {Piacentini}, {Polastri}, {Polenta},
  {Puget}, {Rachen}, {Reinecke}, {Remazeilles}, {Renzi}, {Rocha}, {Rosset},
  {Roudier}, {Rubi{\~n}o-Mart{\'\i}n}, {Ruiz-Granados}, {Salvati}, {Sandri},
  {Savelainen}, {Scott}, {Shellard}, {Sirignano}, {Sirri}, {Spencer},
  {Sunyaev}, {Suur-Uski}, {Tauber}, {Tavagnacco}, {Tenti}, {Toffolatti},
  {Tomasi}, {Trombetti}, {Valenziano}, {Valiviita}, {Van Tent}, {Vibert},
  {Vielva}, {Villa}, {Vittorio}, {Wandelt}, {Wehus}, {White}, {White},
  {Zacchei}, \& {Zonca}}]{planck18}
{Planck Collaboration}, {Aghanim}, N., {Akrami}, Y., {et~al.} 2020, \aap, 641,
  A6, \dodoi{10.1051/0004-6361/201833910}

\bibitem[{{Roming} {et~al.}(2005){Roming}, {Kennedy}, {Mason}, {Nousek}, {Ahr},
  {Bingham}, {Broos}, {Carter}, {Hancock}, {Huckle}, {Hunsberger}, {Kawakami},
  {Killough}, {Koch}, {McLelland}, {Smith}, {Smith}, {Soto}, {Boyd},
  {Breeveld}, {Holland}, {Ivanushkina}, {Pryzby}, {Still}, \&
  {Stock}}]{Roming05}
{Roming}, P. W.~A., {Kennedy}, T.~E., {Mason}, K.~O., {et~al.} 2005, \ssr, 120,
  95, \dodoi{10.1007/s11214-005-5095-4}

\bibitem[{Sanders {et~al.}(2013)Sanders, Soderberg, Foley, Chornock,
  Milisavljevic, Margutti, Drout, Moe, Berger, Brown, {et~al.}}]{sanders13}
Sanders, N.~E., Soderberg, A.~M., Foley, R., {et~al.} 2013, The Astrophysical
  Journal, 769, 39

\bibitem[{Schlafly \& Finkbeiner(2011)}]{galacticrecal}
Schlafly, E.~F., \& Finkbeiner, D.~P. 2011, The Astrophysical Journal, 737, 103

\bibitem[{Schlegel {et~al.}(1998)Schlegel, Finkbeiner, \&
  Davis}]{galacticsurvey}
Schlegel, D.~J., Finkbeiner, D.~P., \& Davis, M. 1998, The Astrophysical
  Journal, 500, 525

\bibitem[{{Smith} {et~al.}(2020){Smith}, {Smartt}, {Young}, {Tonry}, {Denneau},
  {Flewelling}, {Heinze}, {Weiland}, {Stalder}, {Rest}, {Stubbs}, {Anderson},
  {Chen}, {Clark}, {Do}, {F{\"o}rster}, {Fulton}, {Gillanders}, {McBrien},
  {O'Neill}, {Srivastav}, \& {Wright}}]{Smith2020}
{Smith}, K.~W., {Smartt}, S.~J., {Young}, D.~R., {et~al.} 2020, \pasp, 132,
  085002, \dodoi{10.1088/1538-3873/ab936e}

\bibitem[{Smith(2016)}]{Smith2016}
Smith, N. 2016, Interacting Supernovae: Types IIn and Ibn (Cham: Springer
  International Publishing), 1--27, \dodoi{10.1007/978-3-319-20794-0_38-1}

\bibitem[{Speagle(2020)}]{dynesty}
Speagle, J.~S. 2020, Monthly Notices of the Royal Astronomical Society, 493,
  3132

\bibitem[{{STScI Development Team}(2018)}]{synphot}
{STScI Development Team}. 2018, {synphot: Synthetic photometry using Astropy}.
\newblock \doeprint{1811.001}

\bibitem[{Sutherland \& Wheeler(1984)}]{sutherland84}
Sutherland, P.~G., \& Wheeler, J.~C. 1984, The Astrophysical Journal, 280, 282

\bibitem[{Swartz {et~al.}(1995)Swartz, Sutherland, \& Harkness}]{swartz95}
Swartz, D.~A., Sutherland, P.~G., \& Harkness, R.~P. 1995, arXiv preprint
  astro-ph/9501005

\bibitem[{{Tonry} {et~al.}(2019){Tonry}, {Denneau}, {Heinze}, {Weiland},
  {Flewelling}, {Stalder}, {Rest}, {Stubbs}, {Smith}, {Smartt}, {Young},
  {Srivastav}, {McBrien}, {O'Neill}, {Clark}, {Fulton}, {Gillanders}, {Chen},
  \& {Wright}}]{discovery}
{Tonry}, J., {Denneau}, L., {Heinze}, A., {et~al.} 2019, Transient Name Server
  Discovery Report, 2019-1121, 1

\bibitem[{{Tonry} {et~al.}(2018){Tonry}, {Denneau}, {Heinze}, {Stalder},
  {Smith}, {Smartt}, {Stubbs}, {Weiland}, \& {Rest}}]{Tonry2018}
{Tonry}, J.~L., {Denneau}, L., {Heinze}, A.~N., {et~al.} 2018, \pasp, 130,
  064505, \dodoi{10.1088/1538-3873/aabadf}

\bibitem[{Valenti {et~al.}(2007)Valenti, Benetti, Cappellaro, Patat, Mazzali,
  Turatto, Hurley, Maeda, Gal-Yam, Foley, \& Filippenko}]{valenti08}
Valenti, S., Benetti, S., Cappellaro, E., {et~al.} 2007, Monthly Notices of the
  Royal Astronomical Society, 383, 1485,
  \dodoi{10.1111/j.1365-2966.2007.12647.x}

\bibitem[{Valenti {et~al.}(2013)Valenti, Sand, Pastorello, Graham, Howell,
  Parrent, Tomasella, Ochner, Fraser, Benetti, Yuan, Smartt, Maund, Arcavi,
  Gal-Yam, Inserra, \& Young}]{Valenti2013}
Valenti, S., Sand, D., Pastorello, A., {et~al.} 2013, Monthly Notices of the
  Royal Astronomical Society: Letters, 438, L101, \dodoi{10.1093/mnrasl/slt171}

\bibitem[{{Valenti} {et~al.}(2014){Valenti}, {Sand}, {Pastorello}, {Graham},
  {Howell}, {Parrent}, {Tomasella}, {Ochner}, {Fraser}, {Benetti}, {Yuan},
  {Smartt}, {Maund}, {Arcavi}, {Gal-Yam}, {Inserra}, \& {Young}}]{Valenti2014}
{Valenti}, S., {Sand}, D., {Pastorello}, A., {et~al.} 2014, \mnras, 438, L101,
  \dodoi{10.1093/mnrasl/slt171}

\bibitem[{Valenti {et~al.}(2016)Valenti, Howell, Stritzinger, Graham,
  Hosseinzadeh, Arcavi, Bildsten, Jerkstrand, McCully, Pastorello,
  {et~al.}}]{valenti16}
Valenti, S., Howell, D., Stritzinger, M., {et~al.} 2016, Monthly Notices of the
  Royal Astronomical Society, 459, 3939

\bibitem[{{Vallely} {et~al.}(2018){Vallely}, {Prieto}, {Stanek}, {Kochanek},
  {Sukhbold}, {Bersier}, {Brown}, {Chen}, {Dong}, {Falco}, {Berlind},
  {Calkins}, {Koff}, {Kiyota}, {Brimacombe}, {Shappee}, {Holoien}, {Thompson},
  \& {Stritzinger}}]{Vallely2018}
{Vallely}, P.~J., {Prieto}, J.~L., {Stanek}, K.~Z., {et~al.} 2018, \mnras, 475,
  2344, \dodoi{10.1093/mnras/stx3303}

\bibitem[{{Wang} {et~al.}(2021){Wang}, {Lin}, {Zhang}, {Zhang}, {Cai}, {Zhang},
  {Filippenko}, {Graham}, {Maeda}, {Mo}, {Xiang}, {Xi}, {Yan}, {Wang}, {Wang},
  {Kawabata}, \& {Zhai}}]{wang21}
{Wang}, X., {Lin}, W., {Zhang}, J., {et~al.} 2021, \apj, 917, 97,
  \dodoi{10.3847/1538-4357/ac0c17}

\end{thebibliography}
\bibliographystyle{aasjournal}
\end{document}